\documentclass{ws-rv9x6}
\usepackage{amsmath}
\usepackage{graphicx}
\usepackage{ws-rv-van}     % numbered citation/references (default)
\usepackage{wrapfig}
\usepackage{caption}
\def\dd{\mbox{d}}
\def\ve{\varepsilon}

\def\x{{\bf x}}
\def\y{{\bf y}}

\makeindex
\begin{document}

\chapter[First Passage Problems in Biology]{First Passage Problems in Biology}\label{ra_ch1}

\author[T. Chou and M. R. D'Orsogna]{T. Chou\footnote{Depts. of Biomathematics and Mathematics, 
UCLA, Los Angeles, CA 90095-1766, \textrm{tomchou@ucla.edu}} and M. R. D'Orsogna\footnote{Dept. of Mathematics, CalState-Northridge, Los Angeles, CA 91330-8313}}
%\index[aindx]{Author, F.} % or \aindx{Author, F.}
%\index[aindx]{Author, S.} % or \aindx{Author, S.}

%\address{Depts. of Biomathematics and Mathematics \\
%UCLA, Los Angeles, CA 90095-1766\footnote{Affiliation footnote.}}

\begin{abstract}
Applications of first passage times in stochastic processes arise
across a wide range of length and time scales in biological settings.
After an initial technical overview, we survey representative
applications and their corresponding models.  Within models that are
effectively Markovian, we discuss canonical examples of first passage
problems spanning applications to molecular dissociation and
self-assembly, molecular search, transcription and translation,
neuronal spiking, cellular mutation and disease, and organismic
evolution and population dynamics. In this last application, a simple
model for stem-cell aging is presented and some results derived.
Various approximation methods and the physical and mathematical
subtleties that arise in the chosen applications are also discussed.
\end{abstract}
%\markright{Customized Running Head for Odd Page} % default is chapter title.
\body

%\runninglinenumbers

\section{Introduction \& Mathematical Preliminaries}\label{INTRO}

\begin{figure}
\includegraphics[width=4.1in]{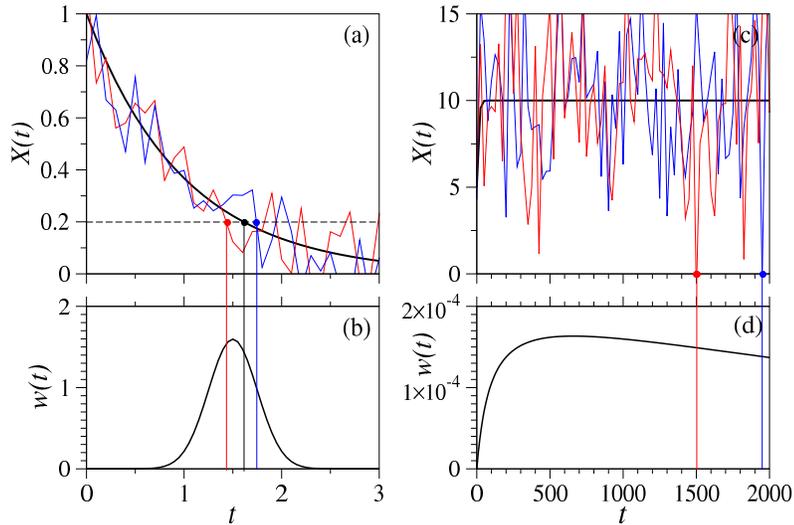}
\caption{Trajectories of a random variable $X(t)$ illustrating typical
  first passage problems.  (a) The deterministic or expected
  trajectory $\langle X(t)\rangle$ (solid black curve) crosses the
  specified threshold $X^{*}=0.2$ at a specific time $T\approx 1.6$;
  however, when fluctuations are explicitly included, the random
  variable $X(t)$ can cross $X=0.2$ at different times $T\approx 1.45$
  and $T\approx 1.7$, as shown by the red and blue trajectories,
  respectively. (b) The distribution of first passage times to $X =
  0.2$. (c) Trajectories corresponding to a birth-death process with
  carrying-capacity (see Eq.~\ref{LOGISTIC} in Section
  \ref{POPULATION}).  In the deterministic model, $X=10$ (the
  carrying-capacity in this example) is a stable fixed point while
  $X=0$ is an unstable one. With an initial condition $X(0) > 0$, the
  deterministic model never becomes extinct $(X^{*}=0)$, but in a
  stochastic model a random (possibly very rare) fluctuation can
  extinguish the system. The distribution of first extinction times is
  schematically shown in (d).}
\label{TRAJECTORY}
\end{figure}

Although mainly studied in physical systems, first passage problems
\cite{REDNER2001} arise in many biological contexts, including
biomolecular kinetics, cellular function, and population dynamics.
First passage problems can be most simply described as finding the
distribution of times according to which a random process first
exceeds a prescribed threshold or reaches a specified configuration,
as described in Fig.~\ref{TRAJECTORY}.  While expectations of moments
of the random variable are often qualitatively captured by using
straightforward approximation methods, other observable quantities
such as first passage times may not be, and stochastic approaches must
be used.

The probability distribution $P(X,t)$ of a stochastic process $X(t)$
may obey a discrete master equation or a Fokker-Planck or Smoluchowski
equation for continuous variables. Other approaches such as the direct
analysis of stochastic differential equations (SDEs) for the random
variable $X(t)$ or analysis of the branching process
\cite{BRANCHING0,HARRIS} describing the evolution of the probability
generating function is also often employed. If the system does not
harbour long-lived metastable configurations, simple mean-field or
closure methods that approximate correlations can be used to
analytically find expected trajectories $\langle X(t)\rangle = \int
XP(X,t)\dd X$ that are often in qualitative agreement with exact
results or trajectories derived from approximate, deterministic
models.

%Within biology, moments of the random variables and first
%passage times are quantities of interest for which theoretical
%estimates are desired.

For example, consider the trajectories depicted in
Fig.~\ref{TRAJECTORY}. Some deterministic trajectories $\langle
X(t)\rangle$ cross a threshold value ($X^{*}=0.2$ in
Fig.~\ref{TRAJECTORY}(a)) at a unique time $T$, which then can be used
as a qualitatively good estimate of the first passage time for the
full stochastic process. However, in other cases, the deterministic
trajectory may never cross a predefined ``absorbing" threshold so that
$T =\infty$.  This is illustrated in Fig.~\ref{TRAJECTORY}(c) where
$X(t)$ never reaches the threshold value $X^{*}=0$.  However, in a
stochastic model, fluctuations can bring $X(t)$ to the threshold
$X^{*}=0$ in finite time. For such cases, there is a clear divergence
between the exit times predicted from a deterministic model
($T=\infty$) and that predicted from a stochastic one ($T< \infty$).

To be more concrete, consider a discrete Markov process for a system
of $N$ states that can be described by the ``forward'' master equation

\begin{equation}
{\partial P_{ki}\over \partial t} = M_{kj}P_{ji}, 
\label{MASTER}
\end{equation}
where $P_{ki}$ is the $N\times N$ matrix of probabilities that the
system is in configuration $k$ at time $t$, given that the system
started in state $i$ at $t=0$. The $N\times N$ transition matrix
composed of transition rates that take state $j$ to state $k$ is
defined by $M_{kj}$. Note that $(k,j)$ indexes all accessible
configurations, including absorbing ones ${\cal A}$ from which
probability density cannot re-emerge. Transition rates out of
configurations ${\cal A}$ are defined to be zero while global
probability conservation requires $\sum_{k=1}^{N}M_{kj}=0$.  As the
dynamics evolve, the flow of probability entering absorbing states
${\cal A}$ cannot exit.  Eventually, the survival probability defined
as $S_{i}(t) \equiv \sum_{k\notin {\cal A}} P_{ki}(t)$ will vanish as
$t\to\infty$. The survival probability $S_{i}(t)$ defines the probability 
that the system has not reached any absorbing configuration 
up to time $t$, given that it started in configuration $i$ at $t=0$.

Since the first passage time distribution can be derived
from $S_{i}(t)$, it is convenient to consider the adjoint equation
that is also obeyed by $P_{ki}(t)$ only if the transition matrix
$M_{kj}$ is time-independent:

\begin{equation}
{\partial P_{ki}\over \partial t} = P_{kj}M_{ji}.
\end{equation}
This ``backward'' equation does not operate on the 
final configurations $k$ so 
one can perform the sum $\sum_{k \notin {\cal A}}$ to 
find an equation for the survival probability 

\begin{equation}
{\partial S_{i}(t) \over \partial t} = S_{j}(t)M_{ji},
% \equiv  -J_{i}(t),
\label{SMS0}
\end{equation}
along with the initial condition $S_{k}(t=0) = 1$ for $k \notin {\cal
  A}$ and ``boundary condition'' $S_{k}(t) = 0$ for $k \in {\cal A}$.

%The last equality in Eq.~\ref{SMS} is simply the time-dependent
%probability flux into the absorbing states given that the system
%started in configuration $i$.  

A physical interpretation of Eq.~\ref{SMS0} can be easily obtained by
considering the {\it lifetime distribution function} which is a sum
over the absorbed states: $F_{i}(t) \equiv \sum_{k\in {\cal A}}
P_{ki}(t) = 1-S_{i}(t)$. We can now identify the rate of change of
$F_{i}$ as the probability flux into the absorbing state ${\cal A}$,
so that $\partial_{t}F_{i} \equiv  J^{\cal A}_{i}(t)$.  Using
$F_{j} \equiv 1-S_{j}$, we can rewrite Eq.~\ref{SMS0} as

\begin{equation}
{\partial S_{i}(t) \over \partial t} = S_{j}(t)M_{ji} = - J_{i}^{\cal A}(t).
\label{SMS}
\end{equation}
The latter is also a statement that the probability of survival
against entering absorbing configurations decreases in time according
to the probability flux into the absorbing states.

From the lifetime distribution $F_{i}(t)$, one can find the
probability that the system reached any absorbing configuration
between time $t$ and $t+\dd t$ as $F_{i}(t+\dd t) - F_{i}(t) =
S_{i}(t) - S_{i}(t+\dd t)$.  Hence, the first passage time
distribution $w_{i}(t)$ can be found from

\begin{equation}
w_{i}(t)\dd t \equiv {\partial F_{i}(t) \over \partial t}\dd t = -{\partial S_{i}(t) \over \partial t}\dd t,
\label{WDSDT}
\end{equation}
allowing calculation of all moments $n$ of the first passage time

\begin{equation}
\langle T_{i}^{n}\rangle = \int_{0}^{\infty} w_{i}(t) t^{n} \dd t.
\label{TN}
\end{equation}
Upon using integration by parts for $n=1$, the mean first passage time is simply 
$\langle T_{i}\rangle = \int_{0}^{\infty} S_{i}(t) \dd t$. Integrating 
Eq.~\ref{SMS} directly, we find an explicit equation for the 
moments of the first passage time into an absorbing state

\begin{equation}
\langle T_{j}^{n}\rangle M_{ji} = -n\langle T^{n-1}_{i}\rangle,
\label{MTN}
\end{equation}
where $\langle T_{j}^{0}\rangle \equiv 1$.  Equations \ref{TN} and
\ref{MTN} have been used to study moments of first exit times for a
random walker to hit either one or two ends of a discrete
one-dimensional lattice \cite{WEISS1981,PURY2003}.

A commonly used approximation to Eq.~\ref{SMS} (see Sections
\ref{RUPTURE} and \ref{SEARCH}) is to assume

\begin{equation}
{\partial S_{i}(t) \over \partial t} \approx -J_{i}^{\cal A}(t)S_{i}(t),
\label{SMFT}
\end{equation}
which is motivated by a mass-action argument of the decay of
probability of being in the initial surviving state $i$.  Here,
$J_{i}^{\cal A}(t)$ is the probability current from state $i$ to
${\cal A}$. However, the RHS of the exact relationship in
Eq.~\ref{SMS} contains the transition matrix $M_{ji}$ which mixes
states $i$ with $j$. Since the approximation in Eq.~\ref{SMFT} does
not resolve the different surviving states, Eq.~\ref{SMFT} is exact
only when there is a single surviving state $i$ that directly
transitions into ${\cal A}$ without any intermediate states.  Another
limit where Eq.~\ref{SMFT} is accurate is if the system mixes quickly
among all surviving states well before being absorbed.  In this case, the
single surviving state $i$ is a lumped average over all the
microscopic states $j$, and first passage can be thought of as slow
degradation of a quasi-steady-state configuration. Equation \ref{SMFT} and the
associated assumptions have been widely used in practice, particularly
to describe bond rupturing in dynamic force spectroscopy of
biomolecules (see Section \ref{RUPTURE}).

Another common representation of stochastic processes that is useful
for modeling biophysical systems is based on continuous variables.
This ``Lagrangian'' representation is particularly suitable for tracking
stochastically-moving, identifiable particles.  Starting from
Eq.~\ref{MASTER}, a continuum formulation can be heuristically
developed by assuming that each configuration is connected to only a
few others.  In this case, indices can be chosen such that the
transition matrix is banded.  For example, a particle at position $i$
on a one-dimensional lattice is allowed to jump only to neighboring
positions $i\pm 1$ with probability proportional to an infinitesimal
increment of time. If the indices label lattice site positions, the
transition matrix will be tridiagonal. Furthermore, if the transition
rates vary slowly from site to site, and the system size $N$ is large,
we can take a continuum limit where the position of a particle $y =
i/N$ and the tridiagonal transition matrix represents a stencil of a
differentiation operator.

Upon defining $P(\{\y_{j}\},t\vert \{\x_{j}\}, 0)$ as the probability
that all particles $j$ are located between $\y_{j}$ and $\y_{j} + \dd
\y_{j}$ at time $t$ given that they were at positions $\{\x_{j}\}$ at
$t=0$, one can Taylor-expand a discrete master equation in a
``diffusion approximation" to find the governing Fokker-Planck or
Smoluchowski equation

\begin{equation}
\begin{array}{rl}
\displaystyle {\partial P(\{\y_{j}\}, t\vert \{\x_{j}\}, 0)\over \partial t} & = 
-\sum_{k=1}^{N} \nabla_{k}\cdot({\bf V}_{k} P) + \sum_{k=1}^{N}\nabla_{k}^{2}(D(\{\y_{k}\})P) \nonumber \\[13pt]
\: & \displaystyle \equiv {\cal L}P(\{\y_{j}\}, t\vert \{\x_{j}\}, 0),
\label{FPE}
\end{array}
\end{equation}
where here, $N$ is the total number of particles, ${\bf V}_{k}$ is the
drift velocity of the $k^{\rm th}$ particle, and the gradient
$\nabla_{k}$ is taken with respect to the coordinates of the $k^{\rm
  th}$ particle.  The density $P(\{\y_{j}\}, t\vert \{\x_{j}\}, 0)$
also obeys the Backward Kolmogorov Equation (BKE) which
is simply

\begin{equation}
\partial_{t} P(\{\y_{j}\}, t\vert \{\x_{j}\}, 0)
= {\cal L}^{\dagger}P(\{\y_{j}\}, t\vert \{\x_{j}\}, 0),
\label{FPADJOINT}
\end{equation}
where ${\cal L}^{\dagger} = \sum_{k}^{N} {\bf V}_{k}\cdot \nabla_{k} 
+ \sum_{k=1}^{N}D(\{\x_{j}\})\nabla_{k}^{2}$
is the operator adjoint of ${\cal L}$. Since ${\cal L}^{\dagger}$ operates on 
the initial positions $\x_{j}$, Eq.~\ref{FPADJOINT} can be integrated 
over coordinates $\y_{j}$ within the domain, excluding the absorbing surfaces. 
The resulting equation for the survival probability analogous to Eq.~\ref{SMS} is 
$\partial_{t} S(\{\x_{j}\}; t) = {\cal L}^{\dagger}S(\{x_{j}\};t)$,
with $S(\{x_{j}\};t=0) = 1$ for all $\x_{j}\neq \partial\Omega_{{\cal A}}$, and 
$S(\forall \x_{j} = \partial\Omega_{\cal A};t) = 0$. From this survival probability,
all moments of the first times any particle hits an absorbing 
boundary $\partial \Omega_{{\cal A}}$ can be derived. Namely, in analogy with 
Eq.~\ref{MTN}, the mean hitting time obeys

\begin{equation}
{\cal L}^{\dagger}\langle T^{n}(\{x_{j}\})\rangle = -n\langle T^{n-1}(\{x_{j}\})\rangle.
\label{LT}
\end{equation}

Both the discrete and continuum stochastic formulations are commonly
applied to physical systems; however, care should be exercised in
using a continuum description as an approximation for a discrete
system where first passage times are sought. Although the continuum
diffusion approximation may be accurate in describing probability
densities of large discrete systems, it often provides a poor
approximation to first passage times of discrete processes.  Indeed,
using a birth-death process with carrying-capacity (see Section
\ref{POPULATION}), Doering, Sagsyan, and Sander \cite{DOERING2005}
show that the effective potential of a discrete system and its
corresponding continuum diffusion approximation differ, leading to
different mean first population extinction times. The discrepancy is
small only when the convective term in the Fokker-Planck equation is
small across all relevant population levels. Thus, depending on the
application, continuum diffusion approximations and their numerical
discretization should be applied judiciously when first passage times
are being analyzed.

The first passage problems defined above assume that one is interested
in the distribution of times of the systems arriving at {\it any}
absorbing configuration. However, there may well be states which are
physically absorbing (into which probability flux enters irreversibly)
but that are not relevant to the biological process.  For example, one
may be interested in the times it takes for a diffusing protein to
first reach a certain target site (see Section \ref{SEARCH} below),
but the protein may degrade before reaching it.  Since decay is
irreversible, the system reaches an ``unintended'' absorbing state
through degradation of the protein.  If one defines ${\cal A}$ to be
only the biologically-relevant absorbing configurations, the
corresponding survival probability $S_{i}(t)$ does not vanish in the
$t\to \infty$ limit because there are other ``irrelevant'' absorbing
states that absorb some of the probability. In other words, if there
are other physical absorbing states competing for probability, the
integrated probability flux $J_{i}^{\cal A}(t)$ into the relevant
absorbing states ${\cal A}$ obeys $\int_{0}^{\infty} J^{\cal
  A}_{i}(t)\dd t < 1$. Also note that since $S_{i}(t\to\infty) > 0$,
the mean first passage time $\langle T_{i}\rangle =
\int_{0}^{\infty}S_{i}(t)\dd t = \infty$. All moments also
diverge. Provided a measurable fraction of trajectories reach the
irrelevant absorbing state, the mean time to arrive at the relevant
absorbing state diverges because these ``wasted'' trajectories will
never reach the relevant states.

A more appropriate measure in cases with ``interfering'' absorbing
states is the distribution of first arrival times {\it conditioned} on
arriving at the relevant absorbing configurations ${\cal A}$. In other
words, we restrict ourselves to the arrival time statistics of only
those trajectories that are {\it not} absorbed by the irrelevant
states. The conditioning is a simple statement of Bayes rule: $J^{\cal
  A}_{i}(t) = J_{i}(t\vert {\cal A}) \times \mbox{Prob(exiting
  through}\, {\cal A})$, where $J^{\cal A}_{i}(t)$ is the overall
probability flux from $i$ into ${\cal A}$, and $J_{i}(t\vert {\cal
  A})$ is the probability flux of annihilation counting those
trajectories that annihilate through the {\it relevant} absorbing
states ${\cal A}$. Since the probability of exiting through ${\cal A}$
is $\int_{0}^{\infty} J_{i}^{\cal A}(t)\dd t$, the conditional first
passage time distribution is

\begin{equation}
J_{i}(t\vert {\cal A})\dd t \equiv 
w_{i}(t\vert {\cal A})\dd t  = {J_{i}^{\cal A}(t) \dd t \over 
\int_{0}^{\infty} J_{i}^{\cal A}(t')\dd t'}.
\label{CONDITIONALW}
\end{equation}
Analogous expressions for the continuum representation 
(Eq.~\ref{FPE}) can be found provided a suitable continuum expression for 
the probability flux is used. As a simple example, consider a single Brownian 
particle with diffusivity $D$ in one dimension with absorbing boundaries at 
$x=\pm 1$. The probability flux through the ends are 

\begin{equation}
\mp D{\partial P(y,t\vert x, 0)\over \partial y}\bigg|_{y=\pm 1} \equiv 
J_{x}(t\vert \pm 1).
\end{equation}
The first passage time distributions sampled over 
only those trajectories that exit, say, $y=+1$ is thus

\begin{equation}
w_{x}(t\vert +1)\dd t = {J_{x}(t\vert +1)\dd t \over 
\int_{0}^{\infty}J_{x}(t'\vert +1)\dd t'},
\end{equation}
which can be explicitly calculated given the solution to the diffusion 
equation (Eq.~\ref{FPE}) for $P(y,t\vert x, 0)$.

The mathematical approaches presented above, along with many
extensions, have been used to model a diverse set of first passage
problems arising in biological systems. In the following sections, we
survey some illustrative examples of such first passage problems that
span length scales ranging from the molecular, to the cellular, to 
that of populations.

\section{Molecular rupture}\label{RUPTURE}

The times over which molecules dissociate play an important role in
chemical biology. For example, ligand-receptor complexes have finite
lifetimes that are important determinants of whether signalling is
initiated. Cell-substrate and cell-cell adhesion are also mediated by
molecules such as glycoproteins \cite{BELL1978}, and knowing the
``strength'' of these macromolecular bonds can reveal insight into the
biological function of macromolecules.

For a simple single-barrier free energy profile, one simple
approximation is to assume a quadratic energy profile and compute the
first passage time distribution to a particular displacement, reducing
the calculation to that of finding the first crossing time of an
over-damped Ornstein-Ulhenbeck process
\cite{NOBILE1985,RICCIARDI1988}.  Another more refined
approximation concatenates two harmonic potentials (one of positive
curvature, one of negative curvature) together to form an approximate
potential. Upon using steepest descents, a simple expression for 
the {\it mean} bond rupturing time starting from the energetic
minimum $\xi_{0}$ can be found in the high barrier (rare crossing)
limit:

\begin{equation}
\langle T(\xi _{0}) \rangle \approx {e^{-(U(\xi^{*}) - U(\xi_{0}))}\over 
2\pi \vert \kappa_{0}\kappa_{*}\vert}.
\label{KRAMERS}
\end{equation}
Here, $\kappa_{0}$ and $\kappa_{*}$ are the curvatures of the
potential at the local minimum and at the top of the barrier,
respectively.  Since the barrier is high, and dissociation is a rare
event, the distribution of rupturing times can be well-approximated by
a single exponential with a dissociation rate $k_{\rm d} \equiv
1/\langle T\rangle$. In addition to the barrier height,
Eq.~\ref{KRAMERS} encodes the shape of the bond potential through the
curvatures $\kappa_{0}$ and $\kappa^{*}$.  However, typical bonds are
sufficiently strong such that their rupture times are too large to be
experimentally accessible. Therefore, bonds are typically pulled by
external forces in ``dynamic force spectroscopy'' (DFS) experiments.

Ideally, in a DFS experiment, the applied force on the bond that is
typically linearly ramped up (in time) until the bond ruptures, and
some properties of the bond trajectories or forces sampled
\cite{EVANS2001,BONDBOOK}. From these data, one may seek to
reconstruct properties of the underlying pre-pulled
potential. Therefore, under an assumption of no rebinding, analysis of
DFS can be reduced to a first passage problem with a time-dependent
potential. Nearly all approaches to this problem have included the
pulling into a time-dependent free energy barrier $U(\xi,t)$, giving
rise to a time-dependent dissociation rate $k_{\rm d}(t)$, which is
then used in the mean-field equation (Eq.~\ref{SMFT}) for the bond
survival probability $\dot{S}(t) \approx - k_{\rm d}(t)S(t)$.  As it
stands, this rate equation does not provide information about the bond
other than the effective barrier height. In order to model finer
effects of the bond energy profiles, shape properties need to be
incorporated into the analysis. The simplest way to do this is to
model how $k_{\rm d}(t)$ depends on the shape of the bond energy,
while still retaining the mean-field assumption (Eq.~\ref{SMFT}) for
the survival probability \cite{BONDBOOK}.

One simple approach is to assume the bond potential contains a barrier
at bond coordinate $\xi^{*}$, beyond which the bond is irreversibly
dissociated.  To approximate the distribution of times for a bond to
spontaneously rupture, one calculates the time it takes for a random
walker to reach the ``absorbing boundary" $\xi^{*}$, given that it
started from an initial position $\xi_{0}$.  The standard calculation
proceeds by solving the Fokker-Planck equation for the probability
density $P(\xi, t\vert \xi_{0}, 0)$ and constructing the corresponding
survival probability $S(\xi_{0}; t) = \int_{0}^{\xi_{*}}P(\xi, t\vert
\xi_{0}, 0)\dd \xi$, or, alternatively, directly solving the Backward
Kolmogorov Equation for $S(\xi_{0};t)$.  The probability density,
survival probability, and rupture time distribution are all easily
solved numerically.  In the over-damped limit of diffusive dynamics,
the {\it mean} bond rupturing time $\langle T\rangle \equiv
\int_{0}^{\infty} S(\xi_{0};t)\dd t$ can be found in exact closed form
for any general free energy profile $U(\xi)$ \cite{GARDINER}.

The simplest way to incorporate a time-varying applied force problem
in the one-dimensional continuum limit is to define an auxiliary time
variable $\tau$ such that $\partial_{t}\tau = 1$. In the backward
equation corresponding to Eq.~\ref{LT}, $\tau$ is an independent
variable \cite{SHILLCOCK1998}

\begin{equation}
\left({\partial \over \partial \tau}+ F(\tau){\partial\over \partial
  \xi} + {\cal L}^{\dagger}\right) \langle T(\xi,\tau)\rangle = -1,
\end{equation}
where $\xi$ is the initial starting coordinate of the bond and 
$F(\tau) = \gamma \tau$ describes a pulling force 
that is increased linearly with rate $\gamma$.
With suitable boundary conditions $\langle T(\xi_{*},\tau)\rangle 
= \langle T(\xi,\infty)\rangle = 0$, one can find 
the expected rupture time $\langle T(\xi,0)\rangle$ numerically. 

Two analytical approximations can be made by assuming the 
pulling force $F$ is fixed. In this case, the solution 
to $\left(F\partial_{\xi} + {\cal L}^{\dagger}\right)
\langle T(\xi,F)\rangle = -1$ is \cite{SHILLCOCK1998}

\begin{equation}
\langle T(\xi,F)\rangle = Q[\exp\left(-U(\xi)+F\xi)\right],
\label{TQ}
\end{equation}
where $Q[...]$ is a complicated, but explicit integral functional \cite{SHILLCOCK1998}.
In a first approximation Shillcock and Seifert \cite{SHILLCOCK1998} 
assumed that the typical rupturing force is determined self-consistently from 
$F^{*} \approx  \gamma \langle T(\xi, F^{*})\rangle$.

A self-consistent approach to estimate the rupture force {\it distribution}
is to solve the mean-field equation $\dot{S}(t;\xi_{0}) = -k_{\rm d}(t)S(t;\xi_{0})$
and use Eq.~\ref{WDSDT} to find

\begin{equation}
w(\xi, t)\dd t = k_{\rm d}(\xi,t)\exp\left[-\int_{0}^{t} k_{\rm d}(\xi,t')\dd t'\right]\dd t,
\end{equation}
where $k_{\rm d}(\xi,t)$ is the time-dependent rate of 
dissociation. Upon using $F(t) = \gamma t$ to 
convert this distribution to a rupture force distribution yields

\begin{equation}
\begin{array}{rl}
w(\xi,F^{*})\dd F^{*} & \displaystyle =
{1\over \gamma}k_{\rm d}(\xi,F^{*})\exp\left[-{1\over \gamma}\int_{0}^{F^{*}}
  k_{\rm d}(\xi,F)\dd F\right]\dd F^{*} \\[13pt]
\: & \displaystyle = {1\over \gamma}k_{\rm d}(\xi,F^{*})\exp\left[-{1\over \gamma}\int_{0}^{F^{*}}
{\dd F \over Q[\exp(-U(\xi)+F\xi)]}\right]\dd F^{*},
\end{array}
\end{equation}
where for the last equality, $k_{\rm d}(\xi,F) \approx 1/\langle
T(\xi,F)\rangle$ and Eq.~\ref{TQ} were used. These and other
mean-field approaches using Eq.~\ref{SMFT} typically lead to a most
probable rupture force $F^{*}$ that is proportional to $\ln \gamma$,
with proportionality factors related to the spatial width and
energetic depth of the underlying bond.  Therefore, the rupture force
distribution measured as a function of loading rate has been widely
used as a quick measure of bond strength.

\begin{figure}
\includegraphics[width=4.4in]{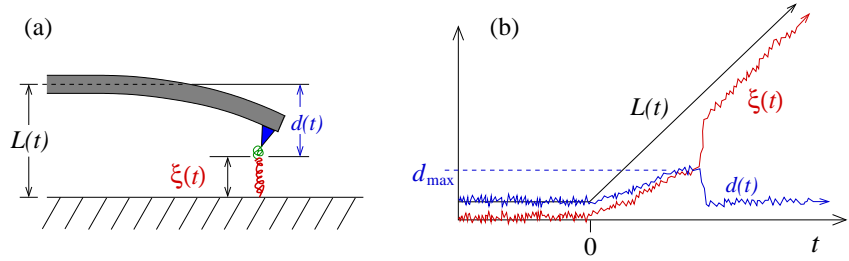}
\caption{Schematic of an AFM-controlled bond-rupturing experiment. The
  AFM is lifted from the rigid substrate with prescribed displacement
  $L(t)$. The deflection $d(t)$ and the bond coordinate $\xi(t)$ sum
  to the total AFM displacement: $L(t) = d(t) + \xi(t)$. The
  deflection $d(t)$ is a measure of the bond-separating force through
  the relation $F(t) = k_{\rm s}d(t)$, where $k_{\rm s}$ is the known spring constant
  of the AFM cantilever.}
\label{DISPLACEMENTS} 
\end{figure} 

While mathematically well-defined, the analyses above neglects a
physical constraint encountered in bond pulling experiments.  As noted
by Qian and others \cite{QIAN1997,QIAN1998,QIAN1999}, the mechanics of
pulling a bond required the introduction of a mechanical spring force,
whether manifested through an atomic force microscope (AFM) tip, an
optical trap, or a pulled magnetic bead. If these devices are pulled
with constant velocity $V$, the actual pulling potential is of the
form $k_{\rm s}(\xi-Vt)^{2}/2$, where $k_{\rm s}$ is the spring constant of the AFM
cantilever. The experimental setup depicted in
Fig.~\ref{DISPLACEMENTS} shows how the pulling force, including an
estimate of the maximum force, can be measured through the deflection
$d(t)$ of the cantilever.

\begin{figure}[t]
%\centering
\includegraphics[width=4.2in]{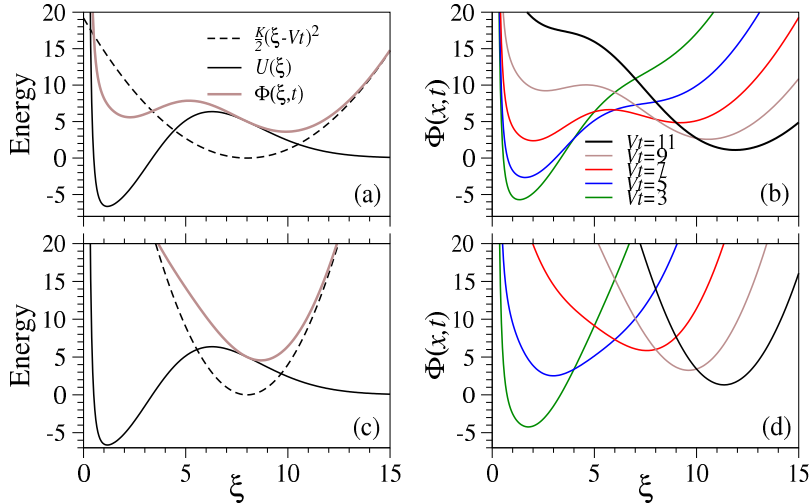}
%\hspace{3mm}\includegraphics[width=1.5in]{FIGS/dists.eps}
\caption{(a) Schematics of free energies as a function of an
  effective, one-dimensional bond coordinate $\xi$. The intrinsic
  molecular potential (black), harmonic potential of the pulling
  device (dashed), and the total potential (brown) are shown. (b) Only
  if the device potential is soft does the total potential
  $\Phi(\xi,t)$ form a barrier as the device is pulled.  (c) If the
  pulling device is stiff, then only a single translating minimum
  arises, (d).  Therefore, only for pulling devices with soft springs
  does a rupture event and corresponding rupture force qualitatively
  arise.}
% Right: The rupture force distribution for various loading
%  rates \cite{MERKEL1999}.}
\label{POTENTIAL}
\end{figure}

This and related approximations are used in combination with specific
bond energy profiles by many authors to derive expressions for rupture
force
distributions\cite{HUMMER2003,DUDKO2003,FUHRMANN2008,GETFERT2009,FREUND2009,FUHRMANN2010,GUPTA2012,GUPTA2013}.
For example, Dudko {\it et al.}\cite{DUDKO2003} treat the ensemble
where the pulling velocity $V$ is specified. They use a mean-field
approximation for the bond survival probability (described in more
detail in Section \ref{SEARCH}) and assume that the total potential is
being shifted at a constant velocity $V$.  For rather general
potentials, they find a mean rupture force $\langle F^{*}\rangle \sim
(\ln V)^{2/3}$, as well as an expression for the rupture force
distribution. These results, however, rely on the use of a soft (small
$k_{\rm s}$) pulling device. As shown in Fig.~\ref{POTENTIAL}, a stiff
puller (large $k_{\rm s}$) results in a single-well effective
potential and a distinct rupture event is precluded
\cite{QIAN1997,QIAN1998,QIAN1999}. In this case, it is not fruitful to
analyze the problem within a first passage time framework, and a more
careful analysis of the force distribution measured during the entire
pulling protocol should be used.

%

%The ultimate goal of using a time-dependent force to generate different
%rupture force distributions is to be able to infer the structure of
%the bond(s) holding molecular components together. All of the models
%used to derive a functional form of the rupture force distribution as
%a function of loading rate $\gamma$ assume a simple potential energy
%profile with few parameters (such as well width and depth). The full
%reconstruction problem of determining a smooth potential energy
%profile from the distribution of rupture times has also been
%considered. 

In general, the problem, as with many inverse problems is
ill-posed \cite{BAL04}.  The reconstruction of a potential from a
single rupture time (or rupture force) distribution starting from a single bond
coordinate is not unique \cite{BAL04}, however, additional
experiments (such as multiple loading forces and multiple starting
bond positions) can give rise to multiple rupture time distributions
that allow for reconstruction of potentials defined by many more
parameters \cite{FOK2010}. The extension of these inverse problems to those using
rupture force distributions derived from different force loading rates
could provide insight into the reconstruction of potentials more
complex than simple harmonic, Lennard-Jones, or Morse type potentials.

\section{Nucleation and self-assembly}\label{SELFASSEMBLY}

\begin{figure}[h]
\begin{center}
\includegraphics[width=4.1in]{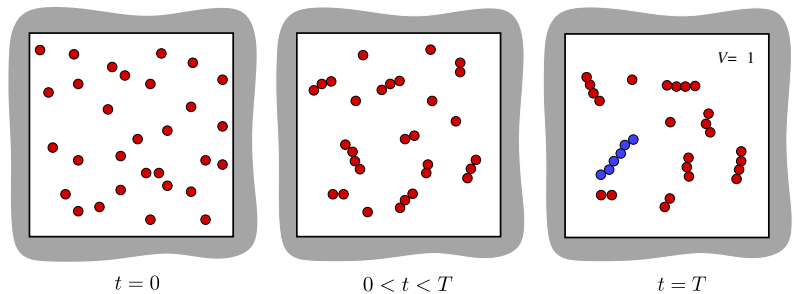}
\caption{Homogeneous nucleation and growth in the slow detachment 
$(q \to 0^{+}$) limit in a closed unit volume
  initiated with $M=30$ monomers. If the constant monomer detachment
  rate $q$ is small, monomers will be nearly exhausted in the long
  time limit.  In this example, we assume that $N=6$ is the maximum cluster size 
and that the first maximum cluster is formed at time $T$ (depicted in blue).}
\label{FIG1}
\end{center}
\end{figure}

A process complementary to dissociation is self-assembly, which also
arises in many biological contexts. The polymerization of actin
filaments \cite{SCIENCE,SEPT2001,MINE2007,KESHET1998,BISHOP1984} and
amyloid fibrils \cite{POWERS2006}, the assembly of virus capsids
\cite{ENDRES2002,ZLOTNICK2007,BRUINSMA} and of antimicrobial peptides
into transmembrane pores \cite{BROGDEN2005,RUTENBERG}, the assembly of
ligands and receptors \cite{DORSOGNA2005,BEL2010}, and the
self-assembly of clathrin-coated pits \cite{CLATHRIN,SHRAIMAN,SENS}
are all important processes at the cellular level that can be cast as
self-assembly problems. Generally, in biological settings, there
exists a maximum cluster size which signals the completion of the
assembly process. For example, virus capsids, clathrin coated pits,
and antimicrobial peptide pores typically consist of $N\sim 100-1000,
N\sim 10-20$, and $N\sim 5-8$ molecular subunits,
respectively. Furthermore, in confined spaces such as cellular
compartments, the total mass is a conserved quantity.
Figure~\ref{FIG1} depicts a homogeneous nucleation process where
monomers spontaneously bind and detach to clusters one at a time.

The classical description of self-assembly or homogeneous nucleation
is a set of mass-action equations (such as the Becker-D\"{o}ring equations)
describing the concentration $c_{k}(t)$ of clusters of each size $k$
at time $t$:

\begin{equation}
\begin{array}{rl}
\dot{c}_{1}(t) & = -pc_{1}^{2}
- pc_{1}\sum_{j=2}^{N-1}c_{j}  + 2 q c_{2} + q\sum_{j=3}^{N}c_{j} \\[13pt]
\dot{c}_{2}(t) & = -pc_{1}c_{2}+ {p\over 2}c_{1}^{2} - q c_{2} + qc_{3} \\[13pt]
\dot{c}_{k}(t) & = -pc_{1}c_{k}  + c_{1}c_{k-1} - q c_{k} + q c_{k+1} \\[13pt] 
\dot{c}_{N}(t) & = pc_{1}c_{N-1} - q c_{N},
\end{array}
\label{HOMOEQN3}
\end{equation}
where for simplicity, we have assumed cluster size-independent 
attachment and detachment rates $p$ and $q$, respectively.
These equations can readily be integrated to 
provide a mean-field approximation to the numbers of 
clusters of each possible size $k$ \cite{DORSOGNA2012}.

Given a total number of monomers $M$ one may be interested in
the time it takes for the system to first assemble a complete cluster
of size $N$. To address such a first passage problem, a stochastic
model for the homogeneous nucleation process must be developed.
Consider an $N$-dimensional probability density $P(n_{1}, n_{2},
\ldots, n_{N};t)$ for the system exhibiting at time $t$, $n_{1}$ free
monomers, $n_{2}$ dimers, $n_{3}$ trimers...and $n_{N}$ completed
clusters. The forward master equation obeyed by $P(n_{1}, n_{2},
\ldots, n_{N};t)$ is \cite{DORSOGNA2012}:

\begin{eqnarray}
\dot{P}(\{n\};t) & = &  -\Lambda(\{n\})P(\{n\};t) +{1\over
  2}(n_{1}+2)(n_{1}+1)W^{+}_{1}W^{+}_{1}W^{-}_{2}P(\{n\};t) \nonumber \\
\: & \: &  \displaystyle  +
\sum_{i=2}^{N-1}(n_{1}+1)(n_{i}+1)W^{+}_{1}W^{+}_{i}W^{-}_{i+1}P(\{n\};t) \nonumber \\
\: & \: & \displaystyle + q(n_{2}+1)W^{+}_{2}W^{-}_{1}W^{-}_{1}P(\{n\};t)  \nonumber \\
\: & \: & \displaystyle + q \sum_{i=3}^{N}(n_{i}+1)W^{-}_{1}W^{-}_{i-1}W^{+}_{i}P(\{n\};t),
\label{MASTERHOMO}
\end{eqnarray}
%\end{widetext}
%
where we have rescaled time to $p^{-1}$.
Here,  $P(\{n\},t) = 0$ if any $n_{i}
< 0$, $\Lambda(\{n\}) = {1\over 2}n_{1}(n_{1}-1) +
\sum_{i=2}^{N-1}\!n_{1}n_{i} + q\sum_{i=2}^{N}n_{i}$ is total rate out
of configuration $\{n\}$, and $W^{\pm}_{j}$ are the unit
raising/lowering operators on the number of clusters of size $j$. For
example,
\begin{equation}
\begin{array}{l}
W^{+}_{1}W^{+}_{i}W^{-}_{i+1}P(\{n\};t)\equiv P(n_{1}+1,\ldots,n_{i}+1,n_{i+1}-1,\ldots;t).
\end{array}
\end{equation}
The process associated with this master equation has been analyzed
using Kinetic Monte-Carlo simulations as well as asymptotic
approximations for the mean cluster numbers in limits of small and
large $q$ \cite{DORSOGNA2012,YVINEC2012}.

\begin{figure}[t]
\begin{center}
\includegraphics[width=3in]{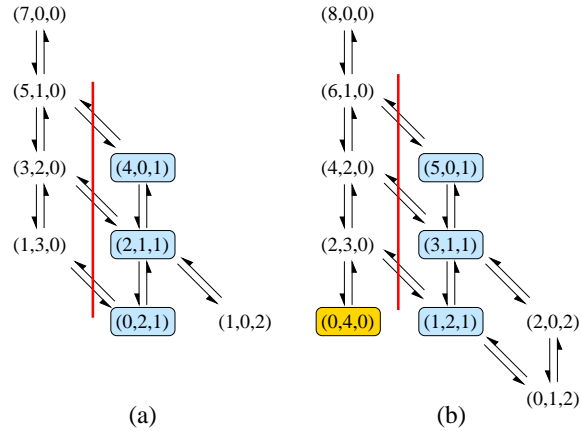}
\caption{Allowed transitions in stochastic self-assembly starting from
  an all-monomer initial condition. In this simple example, the
  maximum cluster size $N=3$.  (a) Allowed transitions for a system
  with $M=7$. Since we are interested in the first maximum cluster
  assembly time, states with $n_{3}=1$ constitute absorbing
  states. The process is stopped once the system crosses the vertical
  red line. (b) Allowable transitions when $M=8$.  Note that if
  monomer detachment is prohibited ($q=0$), the configuration
  $(0,4,0)$ (yellow) is a trapped state. Since a finite number of
  trajectories will arrive at this trapped state and never reach a
  state where $n_{3}=1$, the mean first assembly time $T_{3}(8,0,0)
  \to \infty$ when $q=0$.} 
\label{TREE}
\end{center}
\end{figure}

The first passage problem is to determine the distribution of times
for complete assembly of the largest cluster, $n_{N}=0 \Rightarrow
n_{N}=1$.  For the purpose of illustration, consider a small system
with $M=7$ or $8$, and $N=3$. Since state-space is small, we can
visualize all possible configurations as shown in Fig.~\ref{TREE}. The
first passage time to a maximum cluster, starting from the all-monomer
state ($P(\{n_{i}\};t=0) =
\delta_{n_{1},M}\prod_{i=2}^{N}\delta_{n_{i},0}$) is the time the
system takes to reach any of the states highlighted in blue, to the
right of the red line.

\begin{figure}[t]
%\centering
\includegraphics[width=4in]{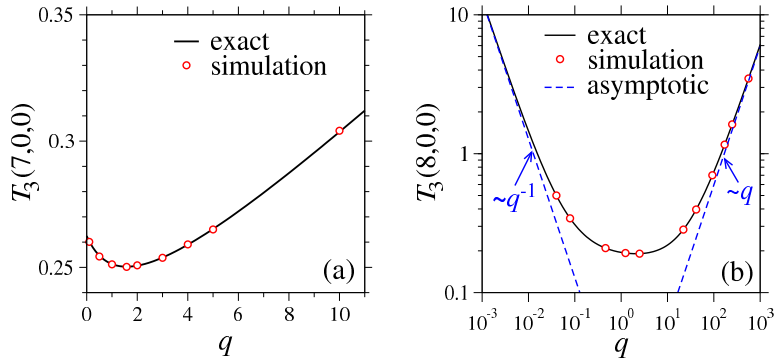}
\caption{Mean first assembly times for $M=7$ and $N=3$ in panel (a)
  and $M=8$ and $N=3$ in panel (b).  The notation $T_{N}(M,0,0)$
  denotes the mean first maximum cluster (of size $N$) assembly time
  $\langle T\rangle$ starting from an initial condition of $M$
  monomers.  Exact results are plotted as black solid lines, while red
  circles are obtained by averaging over $10^5$ KMC simulation
  trajectories.  The dashed blue lines show the $q \to 0$ and $q \to
  \infty$ asymptotic approximations.}
\label{M7}
\end{figure}

In the strong binding limit, when $0 < q \ll 1$ and for $M$ even, one
can find the dominant pathways to a largest cluster and surmise the
leading order behavior $\langle T(q \ll 1) \rangle \sim 1/q$, with a
prefactor that depends nontrivially on $M$ and $N$
\cite{YVINEC2012}. This diverging assembly time arises from trapped
states as highlighted in yellow in Fig.~\ref{TREE}(b). As $q$ is
increased, the likelihood of more paths coming out of the trapped
states is higher, thereby decreasing the expected time to cluster
completion.  Only for the special case of $N=3$ and $M$ odd, where no
such traps exist, is $\langle T(q) \rangle$ a nondivergent ratio of
polynomials in $q$, as illustrated in Fig.~\ref{M7}(a).

In the weak binding, $q \gg 1$, maximum cluster formation is a rare
event and $\langle T(q \gg 1)\rangle \sim q^{N-2}$. Because of these
asymptotic relations, we expect at least a single minimum in the mean
first assembly time as a function of detachment rate $q$ \cite{YVINEC2012}.
Figure~\ref{M7} shows $\langle T(q)\rangle$ as a function of $q$ for
$M=7$ and $M=8$, clearly indicating a shortest expected maximum
cluster formation time at intermediate detachment rates $q$. As long
as $M$ is even or $N\geq 4$, traps states arise and the expected
cluster completion time diverges as $q\to 0$. Thus, in this limit, it
may be physically more meaningful to define the expected assembly time
of a maximum cluster, {\it conditioned} on trajectories yielding
complete clusters.  The above results can also be extended to first
assembly times of the stochastic {\it heterogeneous} nucleation
process \cite{ZHAO2013}.

Ideas of self assembly have also been applied to a structurally more
specific application of linear filament and microtubule growth
\cite{RUBIN1988,BICOUT1996}.  The cell cytoskeleton is a dynamically
growing and shrinking assembly of microtubules and filaments that
regulate cell migration, internal reorganization such as organelle
transport, and mitosis. The assembly and disassembly of microtubules
is a key microscopic process for these vital higher order cell
functions. The molecular players involved in these processes are numerous and
their interaction are biochemically and geometrically complex.
However, one basic feature is that the tips of growing filaments can
exist in a state that promotes elongation, or one that promotes
disassembly.  By switching between these two states, the filament can
be biased to shrink or grow. A first passage problem that has been
studied in this context has been to derive a model for the first
disassembly time of a filament starting from a specific length. Using a
discrete stochastic model describing the probability density for the
number of monomers in a single microtubule, as well as transitions
between growing and shrinking states, Rubin calculated its disassembly time
distribution in terms of modified Bessel's functions \cite{RUBIN1988}.

In later work, Bicout\cite{BICOUT1996} used a semi-Markov model to
describe single filament dynamics. During the growth or shrinking
phases, the length of the filament was assumed to be continuous
variable that increased or decreased according to deterministic
velocities $v_{\pm}$.  However, the switching between growing and
shrinking states was assumed to be Markovian with exponentially
distributed times, with rates $f_{\pm}$.  For this ``Broadwell'' model
we introduce $P_{\pm}(x,t\vert x_{0},0)$ as the probability that the
tip of the filament is moving with velocity $v_{\pm}$ and is located
between position $x$ and $x+\dd x$ at time $t$, given that it was at
position $x_{0}$ at $t=0$. Conservation of probability yields

\begin{equation}
{\partial \over \partial t}\left(\begin{array}{c} P_{+} \\[13pt] P_{-} \end{array}\right)
= {\cal L} \left(\begin{array}{c} P_{+} \\[13pt] P_{-} \end{array}\right),
\label{TELE}
\end{equation}
where 

\begin{equation}
{\cal L} = \left(\begin{array}{cc} -v_{+}{\partial \over \partial x} - f_{+} & f_{-} \\[13pt]
f_{+} & v_{-}{\partial \over \partial x} - f_{-} \end{array}\right),
\end{equation}
which is also known as the ``telegraphers'' equation.  The ballistic
intervals of motion introduces an overall memory into the
dynamics. This can be seen by combining $P_{+}+P_{-} = P$ to find an
equation for the total probability $P(x,t)$ containing terms of the
form $\partial^{2}P/\partial t^{2}$.
 
By using the associated Green's function,
Bicout\cite{BICOUT1996,BICOUT1998} found explicit solutions for the
distribution of lifetimes of a microtubule that started off at a fixed
length $x_{0}$:

\begin{equation}
w(t;x_{0})\dd t \sim t^{3/2}\exp\left[-t/\tau_{\rm c}\right]\dd t.
\end{equation}
The Broadwell model and telegrapher's equation have been used in many
other applications, including gas
kinetics\cite{BROADWELL1964,ILLNER1988} and photon
transport\cite{RUDNICK1997}. In the next section, we
present another example of a first passage problem from molecular
biophysics that involves electron transport and that is also described
by equations similar to Eq.~\ref{TELE}.

\section{Molecular Transport and Search}\label{SEARCH}

A molecular setting in which first passage problems arise in biology
is the so called ``narrow escape problem'', which is simply a higher
dimensional generalization of a high-barrier bond-rupturing problem.  In cellular
environments, numerous confined spaces arise in which molecules
diffuse and react. Typically, a small section of the surface of the
confined space is ``reactive'', {\it i.e.,} contains receptors that
bind diffusing molecules, or is a hole that allows escape into a much
larger volume.  Examples include synaptic clefts connecting neurons,
nuclear envelopes and their associated nuclear pore complexes.

\begin{figure}[ht]
\begin{center}
\includegraphics[width=4.5in]{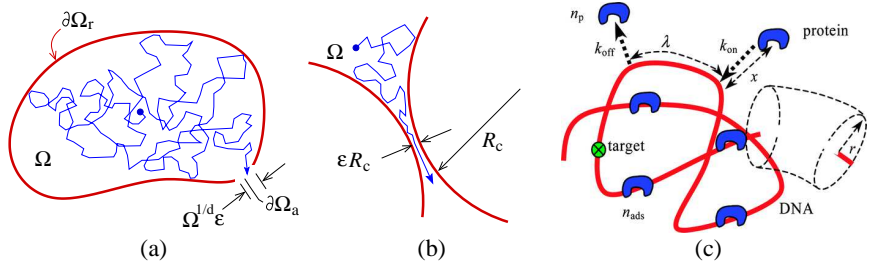}
\end{center} 
\caption{(a) The canonical narrow escape problem. A particle with
  diffusivity $D$ can escape from an asymptotically small
  aperture. The mean time $\langle T(\x)\rangle$ to escape $\Omega$, as
  a function of initial position $\x$ can be calculated in the
  asymptotic limit $\ve \to 0$. (b) An escape problem where the escape
  hatch is at a cusp. (c) DNA target site search problem.  Search is
  facilitated by 1D diffusion along the DNA chain.}
\label{NARROW}
\end{figure}

Mathematically, the problem is described by Fig.~\ref{NARROW}(a) in
which a particle diffuses in the domain $\Omega$, bounded by $\partial
\Omega$. The boundary $\partial \Omega$ is made of two regions, a
reflecting boundary $\partial \Omega_{\rm r}$, and an absorbing one
$\partial \Omega_{\rm a}$, representing a hole or irreversibly binding
surface. Asymptotic results for mean first passage times have been
derived for $\ve = \partial \Omega_{\rm a}/\partial \Omega \ll 1$.
Since escaping is a rare event in this limit, we expect that the
escape time will be insensitive to the starting position.

A number of asymptotic results for the mean escape time of particles
in confined geometries have been determined by Singer, Schuss, and
Holcman \cite{SINGER2008}, as well as Ward, {\it et
  al.}\cite{WARD2010}.

Figure~\ref{NARROW}(a) shows the diffusing particle in a 
volume $\Omega$ that can escape from a small hole of size
$\sim \ve \Omega^{1/d}$, where $d$ is the spatial dimensionality 
of $\Omega$. If $\ve \ll 1$, estimates of the 
mean first exit times have been derived using asymptotic analysis 
of Eq.~\ref{LT} and conformal mapping.
Specifically, in 2D and 3D, for escape from a small hole 
punched through a smooth boundary as shown in Fig.~\ref{NARROW}(a), 
we find

\begin{equation}
\begin{array}{rlr}
\langle T\rangle & \displaystyle \sim {\Omega \over 4\pi D}\left[\log{1\over \ve}
  + O(1)\right], & \mbox{2D} \\[13pt] 
\langle T\rangle & \displaystyle \sim {\Omega^{2/3} \over \ve D}
\left[1+ {\ve \over \pi}\log{1\over \ve}+\ldots\right].
&\quad  \mbox{3D}
\end{array}
\end{equation}
Analogous results were obtained for the constriction escape problem, where
a narrow bottleneck is formed by circles or spheres of radius $R_{\rm c}$
approaching each other or revolved to form a three-dimensional bottleneck:

\begin{equation}
\begin{array}{rlr}
\langle T\rangle & \displaystyle \sim {\pi \Omega \over 2 D\sqrt{\ve}}, & \mbox{2D  } \\[13pt] 
\langle T\rangle & \displaystyle \sim {\Omega\over
  \sqrt{2}R_{\rm c}D}{1\over \ve^{3/2}}, &\quad  \mbox{3D}.
\end{array}
\end{equation}
Similar results have been derived for different geometries such as
diffusion to the tip of a corner, and first passage to the end of a
long neck.  Due to the chosen geometries, escape is a rare event, and
the particle reaches a quasi-steady-state distribution before any
escape has occurred.  Since the time to reach the quasi-steady-state
distribution starting from a specific position is negligible compared
to the mean escape time, all the above results are independent of the
particle's initial position ${\bf x}$.

Another related and biologically important example of first passage is
the search of molecules for their target sites, such as the binding of
transcription factors (sequence-specific DNA-binding proteins) to
their corresponding binding sites along DNA
\cite{HALFORD2004,HU2006,MIRNY2009,VEKSLER2012,VEKSLER2013} (see
Fig.~\ref{NARROW}(c)). These sites are often proximal to the genes
they regulate, although in reality, numerous transcription factors,
including basal factors, RNA polymerase, coactivators, and activators
must assemble before transcription of a specific gene is initiated.
The search problem is of theoretical interest because experimental
search times are much shorter than those estimated from simple 3D
diffusion alone, leading to the idea of facilitated diffusion, a
mechanism whereby more than one transport path is available.  Since
DNA is a linear, often compacted polymer, sections many bases away
from the target may nonetheless be spatially proximal to it. These
physical features have been incorporated into transport models to
estimate the time it takes for an enzyme to bind its intended target
along DNA of arclength $L$.  The original phenomenological model
assumes an effective absorbing sphere of radius $\lambda$ around the
target, where $\lambda$ is the typical contiguous length traveled
along the DNA. A simple heuristic expression for this ``antennae
effect'' on the search time was derived: $\langle T\rangle \approx
(L/\lambda)(\tau_{1}+\tau_{3})$, where $\tau_{1}$ and $\tau_{3}$ are
the typical times spent on the DNA and in the bulk, respectively.  To
obtain realistic search times using this expression requires that the
enzyme spend approximately an equal amount of time on DNA as in the
bulk.  However, in reality, enzymes spend an overwhelming majority of
time associated with DNA. Moreover, this expression breaks down in
certain singular limits such as when the one-dimensional diffusivity
$D_{1} \to 0$, leading to $\tau_{1}\to \infty$.  An improved expression for
the mean search time has been recently derived \cite{CHERSTVY2008},

\begin{equation}
\langle T\rangle \approx {Lr\over 2D_{3} n_{\rm p}}
\left({r \over \lambda n_{\rm ad}} + {\lambda D_{3} n_{\rm p} \over D_{1}
r n_{\rm ads}} + {2D_{3} k_{\rm off} \over k_{\rm on} D_{1}\sqrt{n_{\rm p}}}\right), 
\end{equation}
where $L$ is the arclength of the DNA, $r$ is its effective thickness,
$n_{\rm p}$ and $n_{\rm ads}$ are the number of bulk and adsorbed
proteins, and $k_{\rm on}$ and$k_{\rm off}$ are the attachment and
detachment rates of protein. Note that in this treatment, $k_{\rm on}$
was defined using a reference protein concentration of one molecule
per search volume.  The typical arclength a protein stays within $r$
of the DNA before dissociating is thus estimated to be

\begin{equation}
\lambda \approx {r \sqrt{k_{\rm on} D_{1}}\over \sqrt{k_{\rm off}D_{3} n_{\rm ads}}}.
\label{TSEARCH}
\end{equation}
The result (\ref{TSEARCH}) is able to resolve a number of quantitative
kinetic issues. In particular, Cherstvy, Kolomeisky, and
Korynyshev\cite{CHERSTVY2008} were able to find optimal binding
energies that minimize the search time. Within a realistic
parameter regime, the reduction in search time relative to 3D
diffusion alone can be obtained even for small
$D_{1}/D_{3}$. Additional details and references are found in
Kolomeisky\cite{KOLOMEISKY2011}. Note that all results on this problem 
are independent of the initial starting position of the searching 
enzyme since an initial uniform distribution of enzyme positions is
implicitly assumed.

The molecular search problem is also intimately related to the
filament growth described in the previous section. During mitosis, the
ends of growing and shrinking microtubules emanating from centrosomal
bodies form a party in search of kinetochores that hold together
chromosomes \cite{HOLY1994,WOLLMAN2005}. Using the Green's function
approach of Bicout \cite{BICOUT1996} for a single microtubule as a
starting point, Gopalakrishnan and Govindan \cite{GOVINDAN2011} found
estimates for the search time to one kinetochore

\begin{equation}
\langle T\rangle \approx {e^{\Delta d}\over p}\left(1+ {f_{-}(1-e^{-\Delta d})
\over v_{-}\Delta }\right)\left({v_{+}+v_{-}\over \Delta v_{-}v_{+}} + {1 \over f}\right),
\label{TKINETOCHORE}
\end{equation}
where $\Delta \equiv (v_{-}f_{+}-v_{+}f_{-})/(v_{+}v_{-})$, and
$f$ is the frequency of nucleation of new microtubules from the
centrosome that is located a distance $d$ from the kinetochore target.
The probability that any new microtubule is pointed in the right
direction  and within the capture cone is $p\ll 1$. The microtubule
velocities $v_{\pm}$ and flipping rates $f_{\pm}$ take on the same
meaning as in Eq.~\ref{TELE} used by Bicout to study the lifetime of a
single microtubule. Equation~\ref{TKINETOCHORE} holds only when the
cell radius $R \gg d$.  This and related formulae allow for an easy
determination of optimal parameters that minimize the mean search
time.  The topic of capture of multiple kinetochores associated with multiple
chromosomes has also been treated by Wollman {\it et al.}
\cite{WOLLMAN2005}.

Besides the filament growth and search problems described in Section
\ref{SELFASSEMBLY} and above, two other examples of cellular transport
involving first passage times have been recently discussed: optimal
microtubule transport of virus material to a host cell nucleus
\cite{PLOSONE}, and localization of DNA damage repair enzymes to DNA
lesions \cite{FOK2008,FOK2009}.

When a virus first enters a mammalian host cell its genetic material
needs to be processed and transported into the host cell nucleus
before productive infection can occur.  The transport is often
mediated by molecular motors that carry viral RNA or DNA towards the
nucleus. This process was modeled by a unidirectional convection of
cargo in multiple stages, while detachment of the motor and
degradation of the viral cargo was implemented by a decay term.
Nuclear entry probabilities and conditional first arrival times for
cargo starting at the cell periphery and ending at the nucleus were
calculated \cite{PLOSONE}.  These were found to depend on parameters
describing convection, decay, and transformation in nontrivial ways
which suggested new strategies for drug intervention of the transport
process.

\begin{figure}[ht]
\begin{center}
\includegraphics[width=2.35in]{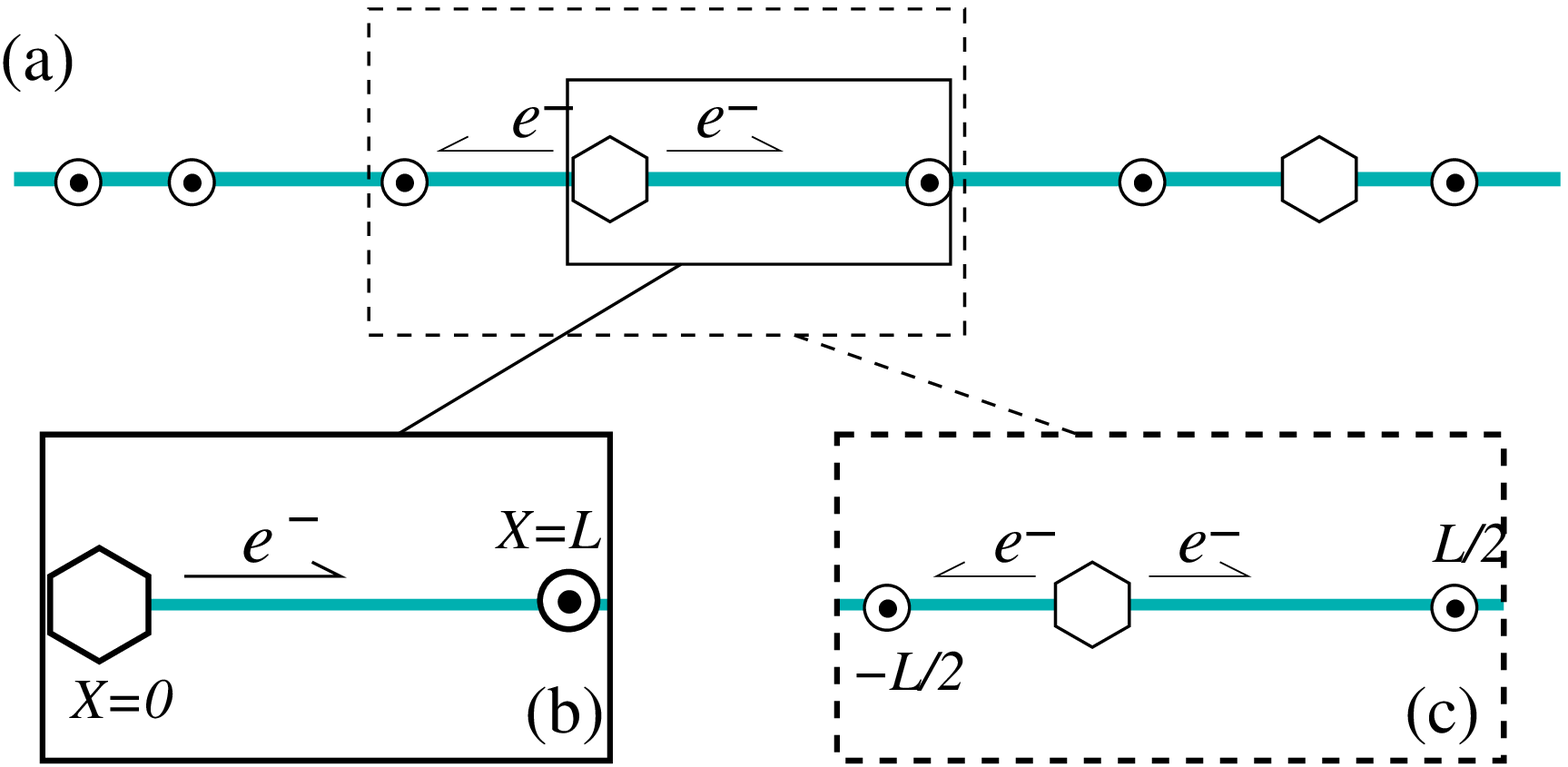}\hspace{3mm}\includegraphics[width=1.95in]{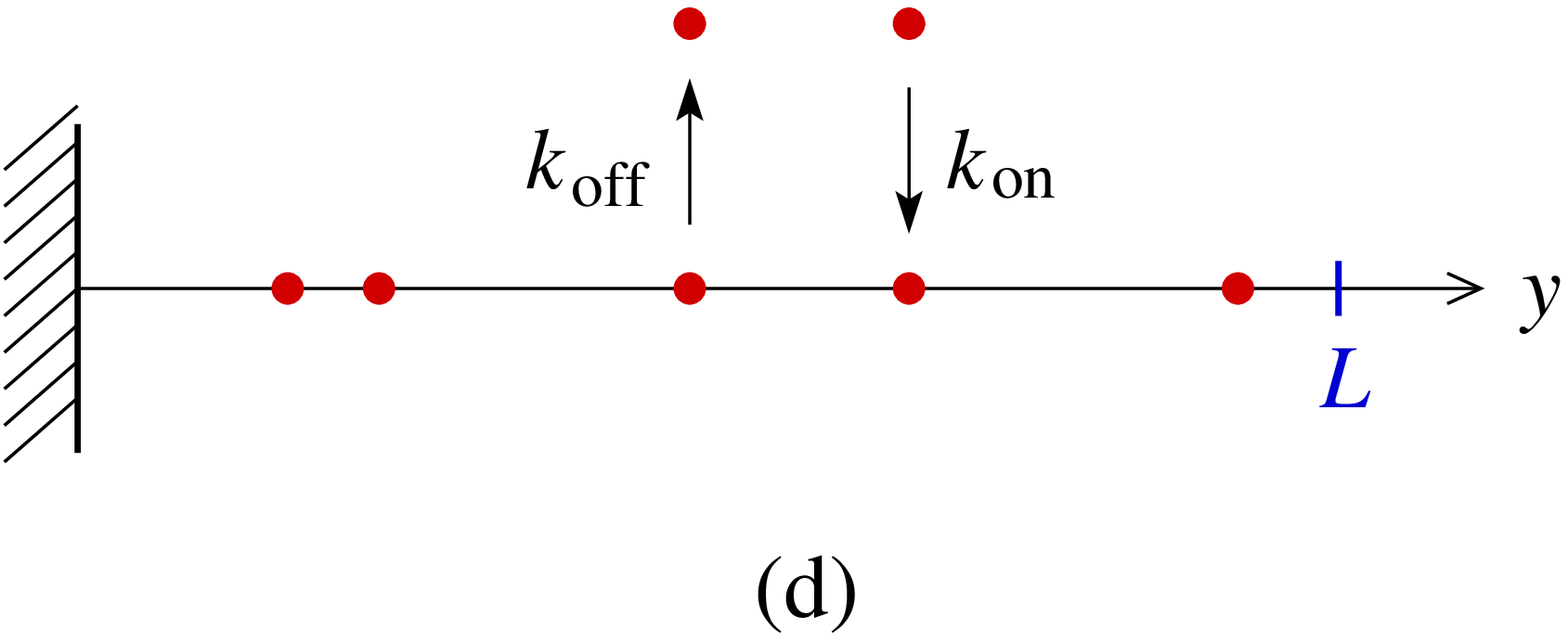}
\end{center} 
\caption{(a) A repair enzyme (hexagon) adsorbs onto a DNA which is
  initially populated by guanine radicals (circled dots) with a
  density $\rho$.  An electron is emitted to the left or right with
  equal probability. The emitted electron has flip rates $f_{\pm}$,
  rightward/leftward velocity $v_{\pm}$ and decay rate $M$.  (b) The
  one-sided Broadwell problem. An electron is emitted from position
  $X=0$ with probability $1$ toward a guanine radical at $X=L$.  (c)
  The two-sided Broadwell problem. An enzyme is deposited between two
  guanine radicals which are at a distance $L$ apart.  Immediately after
  landing inside this segment, an electron is emitted to the left or
  right with equal probability. (d) First passage time to a boundary
  position $y=L$ in the presence of multiple particles undergoing
  Langmuir kinetics.}
\label{BROADWELL}
\end{figure}

Another biophysical example where finding first passage times is
important is the localization of proteins to certain sites on DNA
using an electron ejection mechanism \cite{FOK2008}.  A redox
mechanism for certain DNA repair enzymes to localize near DNA damage
sites has been proposed \cite{BRUNER2000,BARTON2003,ERIKSEN2005}, as
depicted in Fig.~\ref{BROADWELL}(a). Here, a recently deposited repair
enzyme oxidizes by releasing an electron that can either scatter or
absorb at guanine bases and damaged DNA sites. The oxidized repair
enzyme has a higher binding affinity to DNA. However, if the electron
returns, the reduced enzyme will dissociate from the DNA.

Within this overall mechanism, the problems of the first electron
return time, conditioned on its returning arises. The model equation
for this subproblem is identical to Eq.~\ref{TELE} except that $x$,
$v_{\pm}$, and $f_{\pm}$ are the position, speeds, and flip rates of
an electron along the DNA, and decay terms are added to describe the
absorption of electrons ``off'' the DNA. The effective desorption rate
was calculated from the probability and time of electron return. For
repair enzymes that land far from electron absorbing lesions, and if
other electron-absorbing mechanisms are negligible, return of the
emitted electron is likely and the enzyme will detach before it can
diffuse sufficiently far. However, in a finite cell volume, the
detached enzyme reenters the bulk pool and can reattach to the DNA,
potentially closer to the lesion. Deposition near a lesion will likely
be longer-lived because the ejected electron will be more likely
absorbed rather than returning and dissociating the enzyme. In this
way, Fok and Chou \cite{FOK2008,FOK2009} were able to find conditions
under which the repair enzymes statistically localize near
electron-absorbing damage sites on DNA.

Finally, search problems can involve multiple diffusing particles.  In
this case, it is still reasonable to define the state-space in terms
of the positions $\{x_{j}\}, \, 1\leq j \leq N$ for each of, say, $N$
particles. In one-dimension, the first hitting time for any
particle to reach an absorbing point of a line segment has been
examined by Sokolov {\it et al.}\cite{METZLER2005} who considered
noninteracting particles that diffuse and undergo Langmuir kinetics as
shown in Fig.~\ref{BROADWELL}(b). In their study, the authors employ a
mean-field assumption for Eq.~\ref{SMS} where the probability current
$J(t)$ is {\it conditioned} on no other particle having exited the
interval previous to time $t$. The mean-field assumption arises by
expressing this conditioning as $J_{\rm conditioned}(t) =J_{\rm
  unconditioned} S(t)$. The mean-field solution to the probability
$S(t)$ that no particle has hit the target site up to time $t$ is

\begin{equation}
S(t) = J(t)\exp\left[-\int_{0}^{t}J(t')\dd t'\right],
\label{SMFT2}
\end{equation}
where $J(t)$ is the unconditioned probability flux.  Note that for
this approximation to yield physical results, we require

\begin{equation}
\lim_{t\to\infty}tJ(t) > 0
\label{COND}
\end{equation}
in order for $\int_{0}^{t}J(t')\dd t'$
to diverge and $S(t) \to 0$ as $t\to \infty$.
In this problem, the flux was approximated
by $J(t) = -D\partial_{y}n(y,t)\vert_{y=L}$, where $n(y,t)$ is the 
particle density at position $y$ that is found from 

\begin{equation}
{\partial n(x,t)\over \partial t} = D{\partial^{2}n(x,t) \over \partial x^{2}}
-k_{\rm off} n + k_{\rm on},
\label{NEQN}
\end{equation}
where $D$ is the one-dimensional diffusivity, and $k_{\rm on}$ and
$k_{\rm off}$ are the particle adsorption and desorption
rates. Because of the implied infinite bulk reservoir (through rate
$k_{\rm on}$) the mean-field flux satisfies Eq.~\ref{COND}. Even in the
case $k_{\rm off} = k_{\rm on}=0$, if an infinite system size is
assumed, the condition in Eq.~\ref{COND} is also satisfied. In fact, when the
system is infinite, the mean-field assumption in Eq.~\ref{SMFT2} is exact.

A more general approach that does not initially rely on the mean-field
assumption, and can be used for finite-sized systems, is to note that
if the particles are noninteracting, the survival probability $S(t;
\{x_{i}\}) = \prod_{i=1}^{N}S_{1}(t;x_{i})$ is a product of the
survival probabilities of each particle with initial position $x_{i}$.
We assume a finite segment and assume $N$ total of particles,
including those in the bulk.  In this way, we can compute the single
particle probability flux $J_{1}(t) = -D\partial_{y}P_{1}(y,t\vert
x,0)\vert_{y=L}$, and use the exact relation

\begin{equation}
{\partial S_{1}(t;x)\over \partial t} = -J_{1}(t;x) = D\partial_{y}P_{1}(y,t\vert x,0)\vert_{y=L}.
\end{equation}
Using conservation of probability, $\int_{0}^{\infty} J_{1}(t';x)\dd t' = 1$ and, assuming the
initial positions (including the possibility of being detached from the 
lattice) of all particles are identical, we find

\begin{equation}
S(t;x)= \left[1-\int_{0}^{t}J_{1}(t';x)\dd t'\right]^{N}.
\label{SEXACT}
\end{equation}
A direct comparison can be made with 
the mean field result in the case $k_{\rm off} = k_{\rm on} = 0$.
Upon solving Eq.~\ref{NEQN}, we can find the 
the Laplace transform of the single-particle probability flux,
assuming a uniformly distributed initial condition

\begin{equation}
\tilde{J}_{1}(s) = -D{\partial \tilde{n}(x,s)\over \partial x}\bigg|_{x=L} 
= {\tanh L\sqrt{s/D} \over 
L \sqrt{s/D}}.
\label{JTILDE}
\end{equation}
Upon inverse Laplace-transforming, and using the result in
Eq.~\ref{SEXACT}, we can find the exact survival probability. Note
that this result is different from using $NJ_{1}(t)$ for $J(t)$ in the
mean-field approximation Eq.~\ref{SMFT2}. Only in the infinite system
size limit of $L, N\to \infty$, but $N/L = n_{0}$ constant do the
mean-field and exact result $S(t) =
\exp\left[-2n_{0}\sqrt{Dt/\pi}\right]$ coincide. This can be shown
mathematically by using $L = N/n_{0}$ in Eq.~\ref{JTILDE}, inverse
Laplace transforming, substituting the result in Eq.~\ref{SEXACT}, and
taking the $N\to\infty$ limit.  The discrepancy can be most easily
seen by assuming all particles start at $x$ and

\begin{equation}
{\partial S(t;x)\over \partial t} = NS_{1}^{N-1}(t;x){\partial S_{1}(t;x)\over \partial t}
= -NJ_{1}(t;x) S_{1}^{N-1}(t;x).
\end{equation}
For noninteracting particles, the total annihilation flux $J(t;x) = NJ_{1}(t;x)$, and 

\begin{equation}
{\partial S(t;x)\over \partial t} = -J(t;x) S_{1}^{N-1}(t;x) = -J(t;x){S(t;x)\over S_{1}(t;x)}.
\end{equation}
The relative effect of the extra factor $S_{1}(t;x) < 1$ on $S(t;x)$
decreases as $N\to \infty$. It should be stressed that independence of
the diffusing particles allow for the exact analysis above. However,
certain approximate results for interacting particles have also been
obtained \cite{ZILMAN2010}. 

Multiple particle first passage problems also illustrate the concept
of order statistics.  Although Eq.~\ref{SEXACT} provides the survival
probability of a boundary untouched by {\it any} one of the diffusing
particles, one might be interested in the statistics of the first,
second, third, etc., particle to leave the interval, as well as the
complete clearing time distribution.  These order statistics and
asymptotic expressions for the first two moments of the $j$ exit times
have been derived for independent particles diffusing in one-dimension
\cite{YUSTE1996} and $d-$dimensions \cite{YUSTE2001}.

\section{Neuronal Spike Trains}\label{NEURON}

\begin{figure}
\begin{center}
\includegraphics[width=4.2in]{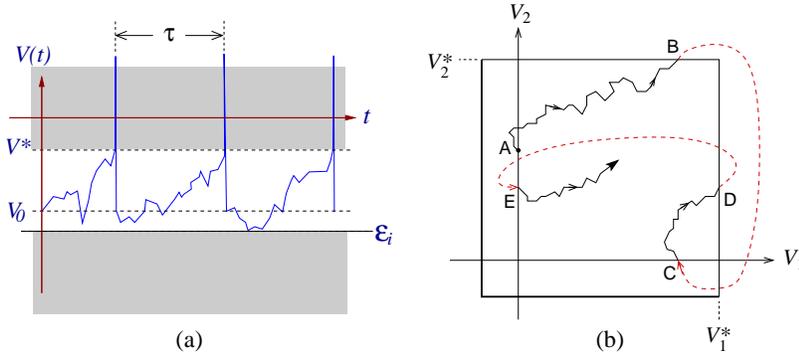}
\end{center}
\caption{First exit times in simple neuronal firing models. (a) A
  schematic time trace of the transmembrane potential showing voltage
  spikes triggered at $V^{*}$ and resetting back to $V_{0}$. The
  subthreshold voltage dynamics is a stochastic processes with the
  interspike time distribution measuring the statistics of the first
  passage time to the threshold voltage. (b) Voltage trajectories for
  two coupled neurons, with transmembrane voltage $V_{1}$ and
  $V_{2}$. If neuron 2 spikes first at point (B), $V_{2}$ spikes and
  quickly resets to point (C). In this example, neuron 1 spikes next
  at point (D), and $V_{1}$ resets to point (E).}
\label{NERVE}
\end{figure}

An important first passage problem within a living, functioning nerve
cell, or group of nerve cells arises in the study of the timing
of electrical spike trains. While modeling the stochastic dynamics of
the membrane potential of a neuron requires taking into account a
large number of detailed microscopic processes, such as nonlinear ion
channel gating and membrane capacitance and leakage, the overall
phenomena of spike trains can be effectively described by a stochastic
process with a threshold membrane potential $V^{*}$.  When the voltage
of a neuron reaches $V^{*}$, highly nonlinear processes take over, the
voltage quickly spikes, and returns to a reset voltage, as shown in
Fig.~\ref{NERVE}(a). The interspike times are distributed
according to the time that the transmembrane potential first reaches
$V^{*}$ after the previous resetting.

A simple one-dimensional stochastic model for predicting interspike
times for a single neuron has been proposed by Stein \cite{STEIN1967}.
Here, the transmembrane voltage is assumed to dissipate through a
``leak'' current, while other connected neurons impart noise to the
neuron of interest. The model implicitly relies on a mean field
assumption in the sense that none of the other neurons are affected by
the behavior of the neuron in question. The ``bath'' neurons provide
random excitatory and inhibitory signals through unspecified
physical connections with the isolated neuron.  Starting from a
stochastic differential equation (SDE) formulation, 
increments of the transmembrane voltage $V$ can be expressed as

\begin{equation}
\dd V = -{V\over \tau} \dd t + a_{\rm e}\dd \pi_{\rm e}(r_{\rm e},t) 
- a_{\rm i}\dd \pi_{\rm i}(r_{\rm i}, t),
\end{equation}
where $a_{\rm e}$ and $a_{\rm i}$ are the fixed amplitudes of the
excitatory and inhibitory spikes feeding into the neuron, and
$\pi_{\rm e}(r_{\rm e},t)$ and $\pi_{\rm i}(r_{\rm i}, t)$ are
possibly time-varying unit excitatory and inhibitory Poisson processes
with rates $r_{\rm e}$ and $r_{\rm i}$, respectively.  Suppose the
voltage starts at $V(t=0) = X$ and that the threshold for spiking is
$V_{*}$. The recursion equations for the moments $M_{n}(X; V_{*})
\equiv \langle T^{n}(X;V_{*})\rangle$ of the interspike times are
\cite{TUCKWELL1980}

\begin{equation}
{X\over \tau} {\dd M_{n} \over \dd X} - r_{\rm e}M_{n}(X+a_{\rm e}) - r_{\rm i}
M_{n}(X-a_{\rm i}) + (r_{\rm e}+r_{\rm i})M_{n}(X) = nM_{n-1}(X), 
\label{DELAY}
\end{equation}
where $M_{0}(X) \equiv 1$. The mean interspike times $M_{1}(X,V_{*})
\equiv \langle T(X)\rangle$ were analyzed by Cope and Tuckwell
\cite{COPE1979} using asymptotic analysis for large negative reset
voltages, and continuing the solutions to the threshold $V_{*}$.
Assuming $a_{\rm e} = a_{\rm i}$, their result for the mean first time
$T(V)$ to spiking starting from an initial voltage $V$ can be
expressed in the form

\begin{equation}
\langle T(X,V_{*})\rangle \approx {1 \over r_{\rm e}}\left[{1\over \tau r_{\rm
      e}}\log\left({V\over a_{\rm e}}\right) + C\left({V_{*}\over
    a_{\rm e}}\right) + \sum_{n=1}^{\infty}A_{n}\left({a_{\rm e} \over
    V}\right)\right],
\end{equation}
where the function $C(V_{*}/a_{\rm e})$ and the coefficients $A_{n}$ 
were numerically found from recursion relations of a set of linear equations.
However, note that the associated equation for the voltage probability density
$P(V,t\vert V_{0}, 0)\dd V$ is

\begin{equation}
{\partial P \over \partial t} = {1\over \tau}{\partial (VP)\over \partial V} + 
r_{\rm e} P(V-a_{\rm e},t) + 
r_{\rm i}P(V+a_{\rm i},t) - (r_{\rm e}+r_{\rm i})P,
\end{equation}
where only arguments of $P$ that are different from $(V,t\vert V_{0}, 0)$ are explicitly 
written. A further simplification can be 
taken by assuming the noise amplitudes $a_{\rm e,i}$ 
are small and Taylor expanding the probability densities
to second order in $a_{\rm e,i}$ (a ``diffusion'' approximation).
The Fokker-Planck or Smoluchowski equation now takes the form

\begin{equation}
{\partial P\over \partial t} = {\partial \over \partial V} 
\left[\left({V\over \tau} - r_{\rm e}a_{\rm e}+r_{\rm i}a_{\rm i}\right)P\right]
+ {1\over 2}\left(r_{\rm e} a_{\rm e}^{2} + r_{\rm i}a_{\rm i}^{2}\right)
{\partial^{2} P \over \partial V^{2}},
\end{equation}
with $P(V,t\vert V_{0}, 0) = \delta(t)$ when $V = V_{*}$.  This model
for subthreshold neuron voltage is simply a first passage
problem of the Ornstein-Ulhenbeck process that has been used to
describe particle escape from a quadratic potential or rupturing
of a harmonic bond. Recasting the problem using a Backward Kolmogorov
Equation, the survival probability (the probability that no spike has
occurred) as well as the moments of the interspike times can be
expressed in terms of special functions \cite{TUCKWELL1980}.  Tuckwell
and Cope \cite{TUCKWELL1980} also provide a careful analysis of the
accuracy of the diffusion approximation in approximating the ``exact''
results from Eq.~\ref{DELAY}.  As expected the diffusion approximation
is accurate in the limit of large excitatory and inhibitory spike noise
rates $r_{\rm e}$ and $r_{\rm i}$, and when the threshold voltage
$V_{*}$ is far from the reset voltage.

Besides simple one-dimensional models, higher dimensional models that
include more mechanistic details of a single neuron have also been
studied. In particular, stochastic first passage problems for
Fitzhugh-Nagumo \cite{TUCKWELL2003} and Hodgkin-Huxley models
\cite{TUCKWELL2005} have been developed. These more complex models
still focus on the voltage dynamics of a single neuron, with the
voltage dynamics of other connected neurons subsumed into the
``noise'' felt by the neuron. Typically, the multiple neuron voltages
can be simultaneously measured using multielectrode recordings,
allowing for the quantification of the correlations between the
spiking times of connected neurons. A first approach for modeling
these higher dimensional data is to treat the stochastic dynamics of a
small number of interacting neurons.  For the two neuron problem
illustrated in Fig.~\ref{NERVE}(b), the dynamics of the subthreshold
voltages of neurons 1 and 2, $V_{1}$ and $V_{2}$, respectively, are
independent of each other, and the probabilities factorize: $P(V_{1},
V_{2})\dd V_{1}\dd V_{2} = P_{1}(V_{1})P_{2}(V_{2})\dd V_{1}\dd
V_{2}$. Interactions between the two neurons occur when either voltage
spikes. A neuron connected to one that spikes can suffer a small
voltage displacement. Rather than treating each neuron as subject to
independent noise, the spiking time statistics of the neurons provide
one component of the random noise of the other neuron. The full
spiking time statistics must be computed
self-consistently. Trajectories in the state space shown in
Fig.~\ref{NERVE}(b) can be described moving along a torus with jumps
in the orthogonal direction each time it crosses circumferentially or
axially. Mathematically, the probability densities for the two
subthreshold voltages obey

\begin{equation}
{\partial P_{i}(V_{i},t) \over \partial t} = {\partial \over \partial
  V_{i}}\left[U_{i}(V_{i})P_{i}\right] + D_{i}{\partial^{2} P_{i}\over
  \partial V_{i}^{2}}, 
\end{equation}
where $D_{i}$ is the voltage diffusivity in neuron $i$.
However, as soon as one $V_{i}$ reaches $V_{i}^{*}$, not only 
does it reset, but $V_{j\neq i} \to V_{j\neq i} + \delta_{j}$
is shifted by $\delta_{j}$.

\section{Cellular and organismic population dynamics}\label{POPULATION}

The simplest nonspatial deterministic population model, describing  growth limitations 
due to a carrying-capacity, centers on the logistic equation

\begin{equation}
{\dd n(t) \over \dd t} = rn(t)\left(1-{n(t) \over K}\right),
\label{LOGISTIC}
\end{equation}
where $n(t)$ is the population density and $K$ is the
carrying-capacity.  This deterministic model has stable fixed points
at $n=0$ and $n=K$.  There are multiple ways to define stochastic
birth-death models that in the mean field limit reduce to
Eq.~\ref{LOGISTIC} \cite{ALLEN}. Nonetheless, all of these models
requires at least one existing organism for proliferation to take
place. Therefore, these models contain an absorbing state at $n=0$,
where the population is extinct. Although the deterministic equation
predicts, at long times, a permanent population $n = K$, a stochastic
model predicts a finite extinction time $T$ after which $n(t\geq T) =
0$. Approximations to this extinction time have been analyzed by
Kessler and Shnerb \cite{KESSLER2007} using a WKB approximation and
Assaf and Meerson \cite{ASSAF2010} using a generating function
approach and properties of the associated Sturm-Liouville equation.
Both methods use the approximation $K \gg 1$, for which extinction is
rare, and a near equilibrium number distribution is first achieved
before an extinction event occurs. This approximation is analogous to
that of assuming ``local thermodynamic equilibrium'' (as opposed to
kinetic theory) for transport calculations \cite{REICHL}. The
probability flux is then constructed from the rate of transport into
an absorbing state from this near equilibrium density.  The
distribution of times for the rare extinction events are nearly
exponential

\begin{equation}
w(t)\dd t \approx \Gamma e^{-\Gamma t}\dd t,
\end{equation}
where to leading order the extinction rate is of the form

\begin{equation}
\Gamma \sim K^{3/2}e^{-K}.
\end{equation}
Note that these results, as with those of the narrow escape problem
(Section \ref{SEARCH}), do not depend on the initial number $n_{0} =
n(t=0)$ because equilibration to a quasi-stationary state occurs on a
time scale much faster than $\Gamma^{-1}$.

Other classic population models, such as models for cell
genotype/phenotype populations, Lotka-Volterra type models
\cite{FREY2012}, and disease models (such as SIS and
SIR)\cite{DYKMAN2008,ARTALEJO2010} have also been extended into the
stochastic realm, and the corresponding exit times into absorbing
configurations analyzed (see Ovaskainen and Meerson\cite{OVA2010} for
a review).  Here, the total organism number is a random variable
determined by the dynamical rules of the model, which may include
``interacting'' effects such as carrying-capacity. The simplest model
for heterogeneity in a birth-death process is the Wright-Fisher model
or, in continuous-time, the Moran model.  The latter is a stochastic
model for two-competing species with numbers $n_{1}$ and $n_{2}$,
where the total population $n_{1}+n_{2}\equiv N$ is fixed. Since
$n_{2} = N-n_{1}$, the problem state-space reduces to one-dimension.
The transition rules in the Moran model are defined by randomly
selecting an individual for annihilation, but instantaneously
replacing it with either one of the same type (so that the system
configuration does not change), or one of the opposite type. The
transition probability in time interval $\dd t$ for converting an
$n_{1}$ individual to an $n_{2}$ individual is thus
$r_{1}n_{1}n_{2}\dd t = r_{1}n_{1}(N-n_{1})\dd t$, while conversion of
$n_{2}$ to $n_{1}$ occurs with probability $r_{2}n_{2}(N-n_{2})\dd
t$. By defining $P(n,t\vert m,0)$ as the probability that there are
$n=n_{1}$ type 1 individuals at time t, given that there were
initially $m$ type 1 individuals, the BKE is simply

\begin{equation}
\begin{array}{rl}
\displaystyle {\partial P(n,t\vert m,0) \over \partial t}
= & \displaystyle m(N-m)\big[r_{1} P(n,t\vert m+1,0) + r_{2}P(n,t\vert m-1,0) \\[13pt]
\: & \hspace{1.3in}- (r_{1}+r_{2}) P(n,t\vert m,0)\big].
\label{MORANKBE}
\end{array}
\end{equation}
Note that $n=0$ and $n=N$ are absorbing states corresponding to the
entire population being fixed to either type 1 or type 2 individuals.
Upon summing $\sum_{n=1}^{N-1}P(n,t\vert m,0) \equiv S(t;m)$, we can
find the corresponding BKE for the probability of survival against
fixation at either $n=0$ or $n=N$. The mean time to fixation can
then be found from inverting the matrix equation 

\begin{equation}
m(N-m)\left[r_{1}\langle T(m+1)\rangle + r_{2}\langle T(m-1)\rangle
-(r_{1}+r_{2})\langle T(m)\rangle\right] = -1,
\end{equation}
with $\langle T(0)\rangle = \langle T(N)\rangle = 0$, to give the well-known result

\begin{equation}
\langle T(m)\rangle = N \sum_{k=1}^{m} {N-m\over N-k} + N \sum_{k=m+1}^{N-1}{m \over k}.
\label{TMORAN}
\end{equation}
If spontaneous mutations are included in the model, there is strictly
no fixation since the states $n=0,N$ are no longer absorbing. Many
generalizations of the Moran model have been investigated, including
extensions to include more species, fluctuating population sizes, and
time-dependent parameters such as the rates $r_{1}(t), r_{2}(t)$
\cite{HOD2003,PERIODIC}.  These extended models are not typically
amenable to closed form solutions such as
Eq.~\ref{TMORAN}. Nonetheless, it is often possible to employ
asymptotic analysis in the large $N$ limit and derive a corresponding
PDE for either the probability density or its generating function.
For example, if one assumes $N\to \infty$ and takes $x = m/N$ one
finds the diffusion approximation for the BKE

\begin{equation}
{\partial S(t;x) \over \partial t} = D_{\rm eff} x(1-x)
{\partial^{2}S(t;x) \over \partial x^{2}}, \quad 0\leq x \leq 1.
\label{MORANCONT}
\end{equation}
Here, we have introduced $D_{\rm eff} = r_{1}N^{2} =
r_{2}N^{2}$. The corresponding PDEs for more complex Moran-type models
are often amenable to analysis, making the Moran model
one of the paradigmatic theories in
population biology and ecology. However, recall from Section
\ref{INTRO} the discrepancy between the first passage times derived
from discrete and corresponding continuum theories \cite{DOERING2005}.
For Eq.~\ref{MORANCONT}, there is no selection or mutation giving rise
to a convection term, so the corresponding mean first passage time
asymptotically approaches the discrete result in Eq.~\ref{TMORAN} as
$N\to \infty$. However, care should be exercised for more complex
models that include effective convection terms.

\begin{figure}
\includegraphics[width=4.5in]{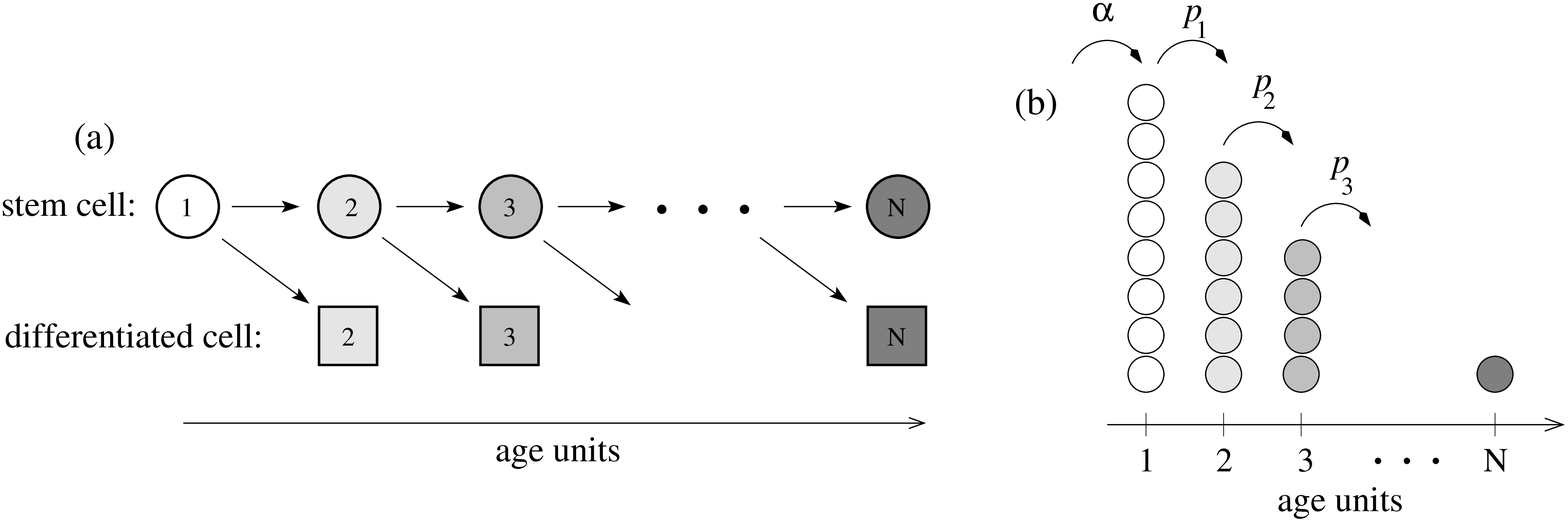}
\caption{Schematic of a reduced model of stem-cell aging. (a)
  Asymmetric division of aging stem-cells. The circles represent stem
  cells, while the squares represent differentiated cells. The
  numerical index represents the age of the cell and is assumed to be
  inversely related to the telomere length. (b) A lattice
  representation of the stem-cell aging model. The rate of asymmetric
  differentiation are shown as $p_{k}$, while the death rates $\mu_{k}$
  at each age $k$ are not indicated.}
\label{AGING}
\end{figure}

Higher dimensional generalizations of these types of discrete models
can also be readily applied to problems in cell population biology
such as cancer modeling and stem-cell proliferation. When the total
population size constraint is relaxed, a linear, multiple state model
shares many mathematical features with the Zero-Range Process (ZRP)
\cite{EVANS2005}, as shown in Fig.~\ref{AGING}. The multiple sites in
such a ZRP might represent the number of cells in a
tissue at a particular mutation stage as the cells progress
towards a cancerous state. Of interest is the first time that a
certain number of cells arrive at the final, ``fully cancerous''
state \footnote{In other contexts, such as individual survival
  probabilities against death from cancer are called Kaplan-Meier
  curves which represent the fraction of a population alive as a
  function of time after the initial diagnosis of cancer}.

Besides multi-hit models of cancer and evolution, the Zero-range
process can also be adapted to model aging in a stem-cell
population. Consider stem-cells that have a limited number of
divisions due to shortening telomeres, ends of their DNA that are
shortened at each division.  Without telomerase to rebuild these ends,
cells will generally be programmed for death. As
shown in Fig.~\ref{AGING}(a), our model assumes that each division
leads to one stem-cell and one differentiated cell, both aged by one
unit (or both with shortened telomeres). Since all cell divisions
are asymmetric, yielding one stem-cell and one differentiated cell, one only
needs to keep track of the number of stem-cells. The forward master
equation for the process has been derived in Shargel, D'Orsogna, and
Chou \cite{SHARGEL2010}, as well as the associated equation for the
generating function:

\begin{eqnarray}
\frac{\partial G}{\partial t} = - \sum_{j=1}^{N-1}(\mu_{j}+p_{j})
z_j {\partial G\over \partial z_j} + \sum_{j=1}^{N}\mu_{j}{\partial G \over \partial z_{j}} + 
 \sum_{j=1}^{N-1} p_{j} z_j{\partial G \over \partial z_{j}} - \mu_{N}
z_N {\partial G \over \partial z_{N}},
\label{GEN}
\end{eqnarray}
where 

\begin{equation}
G(z_1, \cdots, z_N;t) = \sum_{n_j} P(n_1, \cdots, n_N;t) z_1^{n_1} \cdots z_N^{n_N}
\end{equation}
and $P(\{n\};t)$ is the probability that there are exactly 
$n_{k}$ stem-cells of age $k$ at time $t$. If we do not assume
an immigration of new stem-cells defined as having age $k=1$ (as was done in 
Shargel, D'Orsogna, and Chou\cite{SHARGEL2010}), Eq.~\ref{GEN}
can be expressed in the form $\dd G/\dd t = 0$ and solved using the method of
characteristics. The vector of characteristic trajectories 
${\bf Z} = (z_{1}, z_{2}, \ldots, z_{N})^{T}$ can be found
by solving $\dot{{\bf Z}} = {\bf P}{\bf Z} - {\bf M}$, where 

\begin{eqnarray}
\label{matrixeq2}
{\bf P} = \left[ \begin{array}{ccccc}
\mu_1 +p_1 & -p_1 & 0 & \cdots & 0 \\
     0 & \mu_2 +p_2 & -p_2 & \cdots & 0 \\
   0 & \cdots & \mu_j +p_j & -p_{j} & 0 \\
   0 & \cdots & 0 & \mu_{N-1} +p_{N-1} & -p_{N-1} \\
\cdots & \cdots & \cdots & \cdots & \cdots \\
0 & \cdots & 0 & 0 & \mu_{N} 
\end{array} \right]
\end{eqnarray}
and ${\bf M} = (\mu_1, \cdots, \mu_j, \cdots, \mu_N)^{T}$. For an
initial condition of one stem-cell of age $k=1$, these trajectories
can be inverted and expressed in terms of the initial values
$z_{i}(t=0)$, which form the independent variable in the generating
function:

\begin{eqnarray}
\label{generatedisorder}
G({\bf Z};t) &=& z_1 e^{-\Delta_1 t} + \sum_{i=2}^{N} \left[ 
z_i (-1)^{i+1} \left(\prod_{\ell=1}^{i-1} p_{\ell} \right) \sum_{j=1}^{i-1}
\frac{e^{-\Delta_j t}- e^{-\Delta_i t}}{\prod_{k \neq j}^{i} (\Delta_j - \Delta_k)}
\right] + \\ \nonumber \\ \nonumber
&&
1-e^{-\Delta_1 t} + \sum_{i=2}^N \left[  (-1)^{i} 
\left(\prod_{\ell=1}^{i} p_{\ell} \right) \sum_{j=1}^{i-1}
\frac{e^{-\Delta_j t}- e^{-\Delta_i t}}{\prod_{k \neq j}^{i} (\Delta_j - \Delta_k)}
\right], \nonumber
\end{eqnarray}

\noindent
where $\Delta_j \equiv p_j + \mu_j$, for $ 1 \leq j \leq N-1$ and
$\Delta_N = \mu_N$.  From the generating function in
Eq.~\ref{generatedisorder} we can derive the probability 
$P(n_{1}=0,...,n_{j}=1,...,n_{N}=0;t)$ that a
certain age by the descendants of single cell can be found 
at a given age $j$:

\begin{equation}
P(n_1 =1, n_2=0, \cdots, n_N =0)  = e^{-\Delta_1 t},
\label{firstdisorder}
\end{equation}

\noindent while for all other ages $1 < j < N$ we find

\begin{equation}
\begin{array}{rl}
P(0, \cdots, n_{j}=1, \cdots, 0; t) & \displaystyle = (-1)^{j}
\left(\prod_{\ell=1}^{j-1} p_{\ell} \right) \sum_{k=1}^{j-1}
\frac{e^{-\Delta_j t} - e^{-\Delta_{k} t}}{\prod_{i \neq k}^{j}
  (\Delta_k - \Delta_i)} \\[13pt]
\: & = \displaystyle {(pt)^{j-1} \over (j-1)!}e^{-(\mu+p)t},
\end{array}
\label{othersdisorder}
\end{equation}
where the last equality holds in the case where all $p_{i} = p$ and 
$\mu_{i} = \mu$ are age-independent. Finally, the probability for 
complete extinction of the lineage is given by

\begin{equation}
P(\{n\}=0;t) =
1-e^{-\Delta_1 t} +\!\sum_{i=2}^N \left[(-1)^{i} 
\left(\prod_{\ell=1}^{i-1} p_{\ell} \right) \sum_{j=1}^{i-1}
\frac{e^{-\Delta_j t}- e^{-\Delta_i t}}{\prod_{k \neq j}^{i} (\Delta_j - \Delta_k)}
\right].
\label{emptydisorder}
\end{equation}

\noindent
It can be easily verified that the sum of the probabilities in
Eqs.~\ref{firstdisorder}, \ref{othersdisorder} and
\ref{emptydisorder} add to unity, and that  $P(n_1=0,
\cdots, n_N =0; t\to \infty) \to 1$, indicating that a single cell 
will eventually age and that its lineage will go extinct with 
certainty. 

From these probabilities we can construct the probability that the
oldest age reached by a lineage is $Q_k$:

\begin{equation}
Q_{k} = \int_{0}^{\infty}\!\!\left[p_{k-1}P(0,...,n_{k-1}=1,...,0;t)
-p_{k}P(0,...,n_{k}=1,...,0;t)\right]\dd t.
\label{QEQN}
\end{equation}
Equation \ref{QEQN} is derived by 
considering the difference between the probability flux into age $k$ 
and the flux out of age $k$ into age $k+1$ (excluding death). The 
time-integrated result $Q_{k}$ is thus the probability that the 
lineage died at age $k$. For the constant rate case
$p_{i}=p$ and $\mu_{i}=\mu$, we find explicitly

\begin{equation}
Q_{1} = {\mu \over \mu+p}, \quad Q_{k} = {\mu p^{k-1} \over (\mu+p)^{k}}, 
\,\,\,\mbox{and}\,\,\, Q_{N} = {p^{N-1} \over (\mu+p)^{N-1}}.
\end{equation}
From these probabilities, we can define the first passage time to age
$k$ conditioned on the system reaching at least age $k$. Since the
decay at all ages preceding $k$ are ``interfering'' absorbing states,
we can use $J_{1}^{k}(t) = pP(0,...,n_{k-1}=1,...,0;t)$ in
Eq.~\ref{CONDITIONALW} to find

\begin{equation}
w_{1}(t\vert k) \equiv J_{1}(t\vert k) = {(\mu+p)((\mu+p)t)^{k-2}\over 
(k-2)!}e^{-(\mu+p)t},\quad k\geq 2,
\end{equation}
with a corresponding conditional mean arrival time to age $k$:
$\langle T_{1}(k)\rangle = (k-1)/(\mu+p)$. Note that if the decay rate 
$\mu$ is high, the {\it conditional} mean arrival time is small
because only fast trajectories will survive to state $k$.

Our simple stem-cell aging model assumes all divisions are
asymmetric at all ages.  Nonetheless, this model serves as an
illustrative example of an application of a simple Markov process to
cell biology. Indeed, since aging only increases, our model can also
be represented by a simple asymmetric, decaying random walk of a
single ``particle'' in one-dimension, with the position of the
particle representing the age of the single stem-cell in the system at
any given time. The  more complicated approach we have illustrated above allows
our model to be generalized to include effects of multiple initial
stem-cells and symmetric stem-cell division, as well as a more
complete analysis of differentiated cell populations.

\section{Summary}

We have surveyed only a few mathematical and physical models wherein
first passage problems play a central role in the quantitative
understanding of biological observations and experiments. These
applications span all scales from molecular to cellular to
populations. Most applications thus far have been concerned with low
dimensional models with few degrees of freedom.  As measurements improve
and more complex systems can be quantitatively studied, first passage
time problems should become increasingly important in higher
dimensional settings where additional analytic and numerical insights
will be desired. Furthermore, first passage problems provide a new
framework with which to fit experimental data, model biological
processes, and develop inverse problems of model determination.

\section{Acknowledgments}

The authors are grateful to B. van Koten, J. Newby, and J. J. Dong for
incisive discussions and comments.  This work was supported by the NSF
through grants DMS-1021818 (TC) and DMS-1021850 (MD). TC is supported
by the Army Research Office through grant 58386MA.  MD was also
supported by an Army Research Office MURI grant W911NF-11-1-0332.

%\section{Appendix}\index{appendix}
%Appendices should be used only when absolutely necessary. They
%should come before the References.

%\begin{verbatim}
%\begin{appendix}[Optional Appendix Title]
%\section{Sample Appendix}
%Text...
%\begin{equation}
%\mu(n, t) = ...\label{ra_appen1}
%\end{equation}
%\subsection{Sample Subsection}
%Text...
%\begin{equation}
%\zeta\mapsto...\label{ra_appen2}
%\end{equation}
%\end{appendix}
%\end{verbatim}

\bibliographystyle{ws-rv-van}
\bibliography{review12}

%\printindex[aindx]                 % to print author index
%\printindex                         % to print subject index

%%%%%%%%%%%%%%%%%%%%%%%%%%%%%%%%%%%%%%%%%%%%%%%%%%%%%%%%%%%%%%%%%%%%%%%%%%%
\end{document}